**Initial results from a laboratory emulation of weak gravitational lensing measurements**

*S. Seshadri,[1] C. Shapiro,[1] T. Goodsall,[1] J. Fucik,[2] C. Hirata,[2] J.D. Rhodes,[1,2] B.T.P. Rowe,[3] R.M. Smith[2]

[1]Jet Propulsion Laboratory, California Institute of Technology, MS 300-315, 4800 Oak Grove Drive, Pasadena, CA

[2]California Institute of Technology, MC 11-17, 1200 East California Blvd, Pasadena, CA 91125

[3]Department of Physics & Astronomy, University College London, Gower Street, London, WC1E 6BT, UK

Contact information: *suresh.seshadri@jpl.nasa.gov; phone: (818)-354-8370; fax: (818)-393-0045

ABSTRACT

Weak gravitational lensing observations are a key science driver for the NASA Wide Field Infrared Survey Telescope (WFIRST). To validate the performance of the WFIRST infrared detectors, we have performed a laboratory emulation of weak gravitational lensing measurements. Our experiments used a custom precision projector system to image a target mask composed of a grid of pinholes, emulating stellar point sources, onto a 1.7 µm cut-off Teledyne HgCdTe/H2RG detector. We used a 0.88µm LED illumination source and f/22 pupil stop to produce undersampled point spread functions similar to those expected from WFIRST. We also emulated the WFIRST image reconstruction strategy, using the IMage COMbination (IMCOM) algorithm to derive oversampled images from dithered, undersampled input images. We created shear maps for this data and computed shear correlation functions to mimic a real weak lensing analysis. After removing only $2^{nd}$ order polynomial fits to the shear maps, we found that the correlation functions could be reduced to $O(10^{-6})$. This places a conservative upper limit on the detector-induced bias to the correlation function (under our test conditions). This bias is two orders of magnitude lower than the expected weak lensing signal. Restricted to scales relevant to dark energy analyses (sky separations > 0.5 arcmin), the bias is $O(10^{-7})$: comparable to the requirement for future weak lensing missions to avoid biasing cosmological parameter estimates. Our experiment will need to be upgraded and repeated under different configurations to fully characterize the shape measurement performance of WFIRST IR detectors.

# 1. INTRODUCTION

## 1.1. *Science Background*

Gravitational lensing by large-scale structure is an invaluable tool for learning about the dark sector of the Universe. [1,2] Since lensing distortions are caused by the total matter distribution in a structure, they can be used to probe dark matter, which is about 5 times more abundant in the Universe than ordinary matter. Observations of *weak* gravitational lensing are also expected to offer excellent constraints on the properties of dark energy, which accounts for 70% of the cosmic energy density[3] and is believed to be the cause of the observed accelerating expansion of the Universe.[4] Weak lensing refers to very slight distortions to the image of a galaxy, in particular that galaxy's shape and size. These distortions are typically on the order of a few percent, and measuring them in an unbiased way is one of the premier technical challenges for dark energy research.

Weak lensing shape distortions are shown in Figure 1. Shear represents the anisotropic components of the stretching and compression of a galaxy image, for example, turning a circle into an ellipse.

---

[1] Bartelmann, M. and Schneider, P. 2001, Phys. Reports, 340, 4-5, 291
[2] Hoekstra, H. and Jain, B. Annual Review of Nuclear and Particle Systems, 58, 1, 99-123
[3] Planck Collaboration, 2013, arXiv:astro-ph/1303.5062
[4] Albrecht, A., Bernstein, G., Cahn, R., et al. 2006, arXiv:astro-ph/0609591

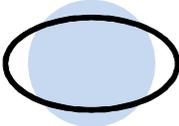

**Figure 1:** Illustration of the shape distortions induced by weak gravitational lensing. The filled circles represent initially circular images. Distorted images are shown as solid lines. Isotropic shape changes are characterized by the convergence, $\kappa$. The anisotropic shear terms, $\gamma_{1,2}$, form a two component pseudo-vector with a magnitude given by $|\gamma|^2=\gamma_1^2+\gamma_2^2$. The two rows show sign conventions for the distortions.

Measuring the ellipticity of a galaxy's light profile therefore provides an estimator for the gravitationally induced shear. The measured ellipticity of a galaxy light profile is the sum of the galaxy's intrinsic shape and the shear:[5]

$$e_i = e_i^{intrinsic} + \frac{1}{m}\gamma_i \qquad \text{Equation 1}$$

where *i* denotes the shear polarization and *m* is a calibration factor of order unity that accounts for shear susceptibility (discussed in 3.5). Since intrinsic ellipticities are not known a priori, weak lensing measurements are necessarily statistical. The intrinsic shapes are approximately randomly oriented; thus

---

[5] For an ellipse with axis lengths a and b, we use the convention that $|e|=|a^2-b^2|/(a^2+b^2)$; see e.g. Rhodes, J., Refregier, A., and Groth, E., 2000, ApJ, 536, 1, 79

$$<e_i> = \frac{1}{m}<\gamma_i> \qquad \text{Equation 2}$$

since the mean intrinsic ellipticity vanishes. Intrinsic ellipticity is typically an order of magnitude larger than the gravitational shear, but a coherent gravitational signal can be measured by averaging over many galaxies to reduce the intrinsic shape noise. With millions of galaxy images, one can construct shear maps that trace the gravitational fields and, therefore, the large-scale distribution of matter in the Universe. Properties of dark matter and dark energy can be constrained by fitting cosmological models to shear statistics such as the shear correlation function:[1]

$$\xi_{ij}(\vec{\theta}) \equiv <\gamma_i(\vec{\theta}_0)\gamma_j(\vec{\theta}_0 + \vec{\theta})> \qquad \text{Equation 3}$$

For a given angular separation θ, this is the mean product of shears of all galaxy pairs separated by θ on the sky.

Weak lensing science is now experiencing a phase of rapid growth. Early measurements of the shear correlation function were achieved at the turn of the 21st century by surveys with small data sets and footprints of, at most, tens of square degrees on the sky.[6] At the time of writing, the most powerful cosmic shear survey has been the Canada-France Hawaii Telescope Lensing Survey (CFHTLenS) with an area of 154 deg² and over 9 million resolved galaxies.[7] Future weak lensing surveys will cover thousands of square degrees and resolve up to billions of galaxies.

---

[6] See Hoekstra, H., Yee, H.K.C., Gladders, M.D. 2002, New Astro. Rev., 46, 12 767 and references therein
[7] Heymans, C., Van Waerbeke, L, Lance, M., et al., 2012, MNRAS, 427, 1, 146

These surveys include the Dark Energy Survey[8], Hyper Suprime-Cam[9], Kilo-Degree Survey[10], the Large Synoptic Survey Telescope[11], the ESA Euclid[12] mission and the NASA Wide Field Infra-Red Survey Telescope (WFIRST) mission.[13]

## 1.2. *Shape Measurement Challenges*

The drastically improved statistical power offered by next-generation surveys is only realized with increasingly strict limits on the allowable galaxy shape measurement errors. For comparison, galaxies used for weak lensing have intrinsic ellipticities of ~0.4 (RMS/component),[14] whereas gravitational shear is $O(10^{-2})$ for a galaxy at z≈1. The correlation between galaxy pairs due to gravity alone is therefore $O(10^{-4})$, while the intrinsic shape noise is $O(10^{-2})$. Future surveys will require that systematic errors in the correlation function be reduced to $O(10^{-7})$ to avoid biasing cosmological parameter estimation.[15] Shape distortions in the data will be corrected using point spread function (PSF) measurements obtained from images of stellar point sources in each exposure. The PSF must be interpolated to the locations of the galaxies and de-convolved from the galaxy images.[16,17] However, the validity of the interpolation and de-convolution will be compromised without sufficient understanding of the various contributions to shape measurement error. These contributions can be split broadly into four categories:

- Optics – PSF distortions due to optical aberrations
- Detectors – generation, collection, and read-out of photo-electric charge by image sensors
- Calibration – converting raw detector data into images that accurately represent the observed sky
- Analysis – extracting sources from calibrated images and creating shear catalogs from them

---

[8] Frieman, J. and Dark Energy Survey Collaboration, 2004, Bulletin of the AAS, 36, 1462
[9] Takada, M. 2010, AIP Conf. Proc., 1279, 120
[10] De Jong, J.T.A., Verdoes, K.G.A., Kuijken, K.H., Valentijn, E.A., Expt'l Astro., 2013, 1-2, 25
[11] Ivezic, Z., Tyson, J.A., Acosta, E., et. al., ArXiv:0805.2366, 2011
[12] Laureijs, R., Amiaux, J., Arduini, S., et al., 2011, arXiv:astro-ph/ 1110.3193
[13] Green, J., Schechter, P., Baltay, C., et al., 2012, arXiv:1208.4012; see also http://wfirst.gsfc.nasa.gov/
[14] See e.g. Mandelbaum, R., Hirata, C.M., Seljak, U., et al., 2005, MNRAS, 361, 4, 1287
[15] Amara, A. and Refregier, A. 2008, MNRAS, 391, 1, 228
[16] Berge, J., Price, S., Amara, A., Rhodes J., 2012, MNRAS, 419, 3, 2356
[17] Gentile, M., Courbin, F., Meylan, G. 2013, A&A, 549, 20

PSF de-convolution itself falls into the analysis category, however it can be made less challenging by attenuating shape measurement errors in the other categories.

Much effort has been invested in using analytical and computational methods to study various shape measurement errors. For instance, the on-going GREAT3 Challenge[18] is investigating the precision of shape measurement algorithms under a worldwide effort (including evaluation of field-dependent effects due to the telescope optics). Prior work examined the impact of modeled detector effects on shape measurements.[19,20,21,22] Current ground and planned space-based missions already have, or are developing, facilities capable of characterizing the fundamental properties of candidate detectors.[23] The data these facilities provide will refine sensor specifications and feed detector physics models that are used in high fidelity, end-to-end *software* simulations of weak lensing observations. The next step in understanding error sources is to perform hardware emulation.

---

[18] Mandelbaum, R., Rowe, B., Bosch, J., et al. 2013, arXiv:1308.4982
[19] Rhodes, J., Leauthaud, A., Stoughton, C., et al. 2010, PASP, 122, 439439
[20] Rhodes, J., et al. 2013, in prep.
[21] Cropper, M., Hoekstra, H., Kitching, T., et al. 2013, MNRAS, 431, 4, 3103-3126
[22] Massey, R., Hoekstra, H., Kitching, T., et al. 2013, MNRAS, 429, 1, 661

## 1.3. *The Precision Projector Lab*

The Jet Propulsion Laboratory (JPL) and Caltech Optical Observatories (COO) have jointly developed the Precision Projector Lab to perform end-to-end *experimental* emulation of weak lensing observations in the lab. We have built a system that projects precisely controlled images onto CCD, CMOS or IR detectors. These images can emulate astronomical objects, such as stars and galaxies, or spectra. We are presently using this system to assess the impact of near-infrared (NIR) detectors, such as those proposed for WFIRST, on weak lensing shape measurement. These detectors have several known non-idealities whose impact on shapes is poorly understood. These include:

- inter-pixel capacitance (IPC)[24,25] – electronic cross-talk between pixels during read-out
- non-linearity[26] -- pixel values not proportional to number of incident photons
- persistence[27] -- exposures contain artifacts from previous exposures
- reciprocity failure[28] -- flux-dependent quantum efficiency

---

[23] Projects with detector test facilities: WFIRST, LSST and Euclid
[24] Seshadri, S., Cole, D.M., Hancock, B.R., et al., 2008, Proc. SPIE, 7021, 11
[25] Moore, A.C., Ninkov, N. and Forrest, B., 2006, Proc. SPIE, 5167
[26] Bezawada, N., Ives, D. and Atkinson, D., 2007, Proc. SPIE, 6690, 669005
[27] R.M. Smith, Zavodny, M., Rahmer, G., et. al. 2008, Proc. SPIE, 7021, 70210K
[28] Bohlin, R., Linder, D., and Riess, A., 2005, "NICMOS Instrument Science Report 2005-002," Space Telescope Science Institute, Baltimore, Md.

In addition, while the intra-pixel response (the dependence of the pixel value on relative position of an image within a pixel) of these detectors has been measured,[29,30] there is little data on their variability, or the errors in pixel boundary locations which are likely to cause greater distortions in undersampled images, such as those expected from WFIRST.

The ultimate goals of our emulation experiments are to validate candidate sensors under realistic use scenarios to determine (i) the scale of detector-induced errors under planned observing conditions – particularly in undersampled images; (ii) whether proposed calibration and analysis methods for shape measurement will deliver the required mission performance when real detectors and optics are used and (iii) the impact of design choices such as plate scale (a major cost driver) on galaxy shape measurement accuracy. We also plan to assess the relative importance of different detector effects on PSF calibration and shape measurements. While our focus is on the requirements for the WFIRST mission, our methodology is general and may ultimately be useful to other weak lensing projects. Our results should also be instructive in the design of realistic software simulations of weak lensing observations currently being developed to test shape measurement methods.[18]

---

## 1.4. *Scope of this Paper*

In this paper we present preliminary results from our characterization of a Teledyne HgCdTe Hawaii-2RG detector. In particular, we measure a shear correlation function on projected point sources and find that it is encouragingly small. We do not attempt to isolate or model specific contributions to the measurement at this stage, but we place an $O(10^{-6})$ upper limit on the spurious contribution that the detector would produce in a weak lensing analysis.

In section 2, we describe the projector system hardware, discuss its emulation capabilities and summarize test conditions for the data reported in this paper. In section 3, we summarize the calibration and analysis pipeline used to process the raw image data, including our methods for oversampled image reconstruction, shape measurement, and correlation function computation. In section 4, we present data quantifying system performance. In section 5, we present results of the correlation function calculated from ellipticity maps of point source images. We conclude in section 6 with a discussion of the relevance of our measurements for future weak lensing science and a summary of our strategy for the next phase of our investigation.

## 2. EXPERIMENTAL SETUP

## 2.1. *System Hardware Description*

The Precision Projector was designed to support laboratory, end-to-end, emulation of astronomical observations from space to demonstrate feasibility of shape measurements such as gravitational weak lensing and high precision photometry using candidate image sensors. This emulation capability is motivated by the widespread experience that numerical modeling informed by detector characterization has frequently proven to be an inadequate predictor of in-flight performance. Common reasons are that typical detector characterization uses flat fields and dark exposures rather than realistic scene projection, and often do not use the planned readout waveforms and observing cadences. The process of experiment emulation also enables validation and optimization of essential calibration procedures such as the dithering and image recombination methods used for undersampled imaging.

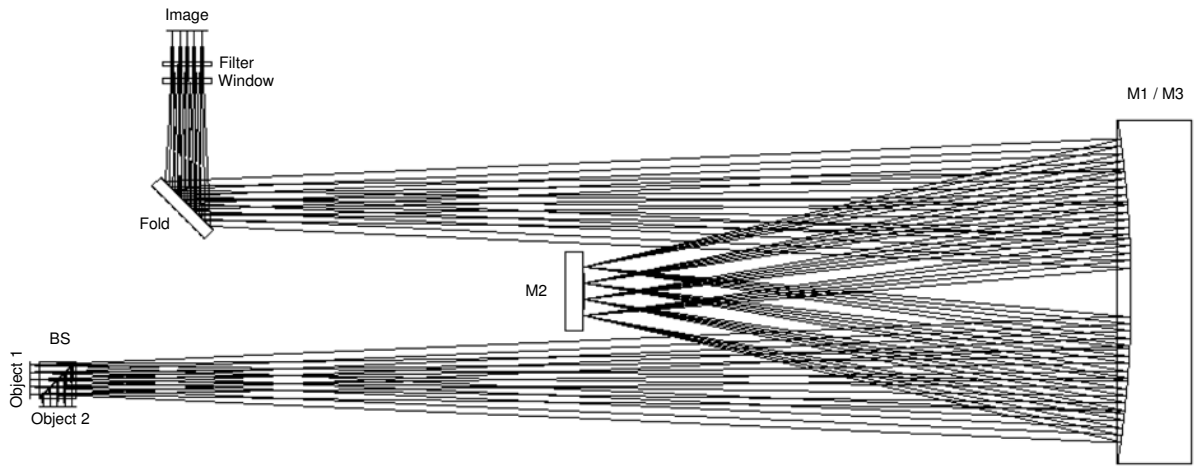

Figure 2: Ray trace of the 1:1 Offner-based re-imaging system. Light from a back-illuminated target mask located at Object 1 is passed through spherical primary (two-reflections) and secondary mirrors (with λ/100 surface figure) to a cryogenically-cooled detector via a fold mirror, window and cryogenic long-wave blocking filter.

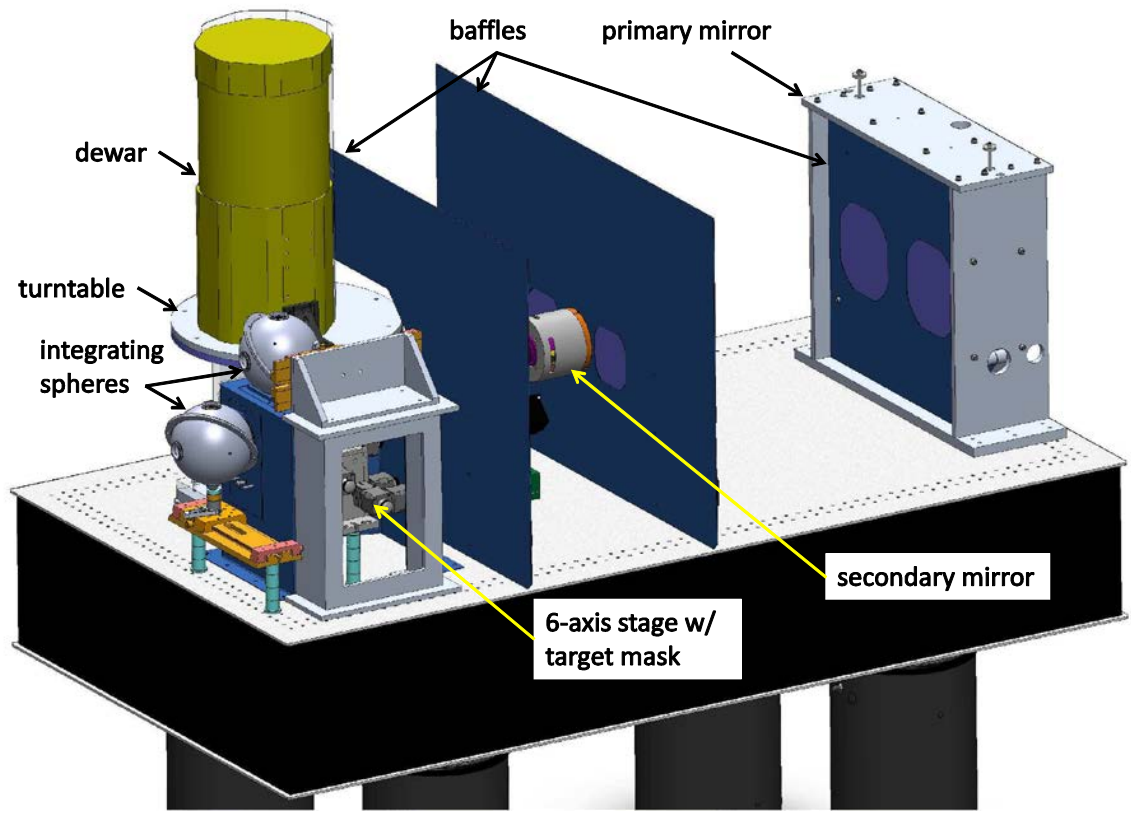

Figure 3: 3-D CAD model of the system. Dual integrating spheres and a beam splitter allow for flat illumination without mask removal. A 6-axis flexure-stage supports 50 mm x 50 mm target masks, which it can translate by 4 mm and rotate up to ±3° in any direction. The rotatable pupil stop at the secondary mirror supports both circular and elliptical stops with variable focal ratio (f/8 or larger). A cryogenic dewar is mounted onto a stage that enables the detector to rotate 360° relative to the optical axis. A thermally isolating enclosure (not shown) and internal baffles, minimize stray light and temperature variations.

Figure 2 and Figure 3 illustrate optical and 3D Computer-Aided Design (CAD) models of the system. A moderately large Offner Relay projects an unmagnified image of a high (50nm) resolution chrome-on-glass lithographic mask onto the detector under test with very low optical aberration and extremely low image distortion over the full UV to NIR wavelength range. By making optical sources of ellipticity significantly smaller than those found in astronomical telescopes, other sources of ellipticity originating in the detector are measured with greater certainty. This all-reflective (thus achromatic) design was developed in the 1970's for high resolution imaging of semiconductor masks.[31] The small residual optical aberration is astigmatism, which unfortunately makes ellipticity a strong function of focus, but this effect is still weak compared to other optical designs considered. The 750mm focal length was chosen to greatly exceed the width of the detector so that contribution of the optical design to image ellipticity (<0.1%) varies only weakly across the field of view and is typically smaller than the ellipticity due to optical misalignment. The baffles and optics deliver optimum performance over the full 40mm square field at f/11, but vignette beyond 20 mm at f/8.

Readily interchangeable object masks each contain many thousands of circular or elliptical apertures. These are a few pixels across to represent extended sources, or have a 3μm diameter to approximate point sources. The resolved apertures are small enough that, when convolved with the PSF, they deliver profiles at the detector which approximate weak lensing targets. Most source galaxies in a large weak lensing survey such as WFIRST will only be barely resolved, i.e. the widths of their light profiles will be comparable to the width of the PSF. For example, WFIRST, LSST and Euclid forecasts assume galaxy widths of at least 0.8 times the PSF size.[32] All experiments reported in this paper use point sources—unresolved 3μm open apertures with flat-topped intensity profiles.

---

[31] A. Offner 1975, Opt. Eng. 14, 2, 130
[32] see Spergel, D., Gehrels, N., Breckinridge, J., et al., 2013, "Wide-Field InfraRed Survey Telescope-Astrophysics Focused Telescope Assets WFIRST-AFTA Final Report," arXiv: 1305.5422

Motorized six-axis motion control of the object mask provides fine adjustments to tip, tilt and focus, in-plane rotation angle, and XY translation (1 μm resolution, 1/18$^{th}$ pixel) for "image dithering". Mask illumination is provided by a high power, thermally regulated 880nm LED with precision current regulation, via an integrating sphere located near the mask surface. A fused silica beam-combining cube provides flat field illumination from a second integrating sphere when required for calibration or for adding a stable, flat background illumination.

Table 1: Summary of capabilities of the Precision Projector system.

| | |
|---|---|
| **Focal length** | 75 cm |
| **Focal ratio (f/#)** | variable (8 – 44) |
| **Illumination wavelength** | Nominal range: 0.35 - 2.0 μm |
| | Limited by atmospheric absorption at short wavelengths and thermal background at long wavelengths |
| **Optimal field of view** | 40 x 40 mm$^2$ at f/11 or larger; 20 x 20 mm$^2$ at f/8 |
| **Target mask** | 50 x 50 mm$^2$ |
| | About a dozen patterns have been made to date, with unit cells that are a combination of point sources and extended objects of differing sizes and orientations. Typically 5,000 to 10,000 objects per image. The mask used in the experiments described in this paper contained 1884 spots per image. |
| **Target mask positioning** | Six-axis stage:<br>• 4 mm translation with 1 μm step<br>• ±3$^o$ rotation with 1.5 arcsec step |
| **Detector** | Accepts cryostats for large format detectors (NIR, CCDs, CMOS) with cryogenic band pass filter |
| **Detector positioning** | Rotational stage with detents at 15$^o$ increments |

A variety of interchangeable circular and elliptical pupil stops, placed within ~100μm of the surface of the secondary mirror, provide a range of PSF widths with predictable profiles since they are strongly dominated by diffraction. Care has been taken to baffle the light path to avoid scattered light and ghost reflections, which could perturb the perceived PSF shape.

System capabilities are summarized in Table 1, with full details given elsewhere.[33]

## 2.2. *Emulation Capabilities and Sampling Scale*

The projector's PSF size and shape are set almost entirely by diffraction and are, thus, well controlled by changing the pupil stop and wavelength. The focal ratios supported by the projector span those likely to be chosen for surveys from space, allowing experiments to show how performance scales with PSF sizes of interest. An important advantage of our PSF control is the ability to properly emulate images over a range of sampling regimes, from over-sampled to highly aliased. Left unmitigated, aliasing causes shape distortions and astrometric errors. The sampling quality of an imaging system can be quantified by the sampling factor

$$Q = \frac{u_{\text{samp}}}{u_{\text{max}}}$$  Equation 4

where $u_{\text{max}}$ is the maximum spatial frequency (band limit) admitted by the optical system and $u_{\text{samp}}$ is the sampling frequency of the detector. Both spatial frequencies are measured in cycles/µm at the focal plane.

The spatial band limit of an optical system projecting a monochromatic image at the focal plane is $u_{\text{max}} = 1/(f\lambda)$, where f is the focal ratio (f/#) and $\lambda$ is the illumination wavelength. The sampling rate of the detector is $u_{\text{samp}} = 1/P$, where P is the pixel pitch. The sampling factor is therefore given by

$$Q = \frac{f\lambda}{P} \qquad \text{Equation 5}$$

---

[33] Smith, R.M., Fucik, J. et al., in prep.

Since the full-width at half-maximum (FWHM) of the PSF of an Airy spot is ~1.03*fλ, the sampling factor of a diffraction-limited system is approximately the ratio of the FWHM to the pixel size.

When Q ≥ 2, the image contains no aliasing and is in the over-sampled regime. Q=2 corresponds to "critical" sampling at the Nyquist frequency, which is twice the spatial bandwidth of the system. For 1 < Q < 2, some spatial frequencies in the image have been aliased, resulting in the images missing some high-frequency information; this is called "weak undersampling". When Q ≤ 1, every frequency in the image is aliased; this is called "strong undersampling". Note that the sampling factor characterizes the diffraction of light at the pupil and is independent of the target mask. Thus, one cannot avoid aliasing of objects on the mask by simply enlarging them. One can, however, attenuate high frequency spatial structure on the target mask by blurring (e.g. with a Gaussian filter).

Operating at low Q values with a narrow PSF allows us to test our ability to reconstruct oversampled data from multiple dithered, undersampled images. Ideally, the number of dithers needed increases approximately as $(2/Q)^2$. Thus, a 2x2 grid of dithers spaced precisely 1/2 a pixel apart can be used to reconstruct Nyquist-sampled or oversampled images when Q=1. In practice, precise dithers cannot be obtained. For instance, the WFIRST survey strategy will image each source using multiple detectors in the focal plane to overcome the loss of area due to gaps between individual imagers. The scan pattern, combined with the varying plate distortion across detectors, will yield effectively random relative dither locations (modulo one pixel) for a given source. With randomized relative positions, more dithers are needed to ensure they contain enough information to achieve Nyquist-sampling. The Precision Projector system implements a semi-automated random dithering mode on the mask scan stage to mimic this feature of the observations.

## 2.3. *Test Conditions*

All experimental data described in this paper were collected using a 2k x 2k format Teledyne H2RG HgCdTe detector with a cut-off wavelength of 1.7 µm. This detector is similar to the 2.4 µm cut-off version base-lined for the WFIRST Design Reference Mission #1 (DRM1) but has much lower sensitivity to thermal background[34] and much lower thermal dark current signal at any given temperature. The WFIRST Astrophysics Focused Telescope Assets (AFTA) DRM would use 4k x 4k H4RG detectors from the same family.[35] Both integrating spheres in the projector were equipped with high-power LED lamp illumination sources centered at 880 nm with 20 nm pass band. A z-band filter centered at 877.5 nm with a 105 nm bandwidth was mounted inside the cryostat to provide rejection of background thermal radiation. This filter has an extinction coefficient >$10^{-4}$ out of band over the response spectrum of the detector. WFIRST will image in a band centered at 1.6µm, however, we use an 880nm LED source for several reasons: (i) at this wavelength it is relatively simple to setup a silicon photodiode feedback loop to maintain stable

---

[34] Blank, R., Anglin, S., Beletic, J. W., et al. 2012, Proc. SPIE 8453, 845310
[35] There are three WFIRST DRMs: see Green et al. [Ref. 12]; See also Spergel, D., Gehrels, N., Breckinridge, J., et al., 2013, "Wide-Field InfraRed Survey Telescope-Astrophysics Focused Telescope Assets WFIRST-AFTA Final Report," arXiv: 1305.5422

lamp intensity; (ii) high-power LEDs are readily available; (iii) thermal background radiation can be easily suppressed with a cold filter. We will however adjust the operating wavelength to investigate possible wavelength-dependent detector effects (e.g. sub-pixel response variations) in the near future.

In practice, the sampling factor Q (see section 2.2) is the most relevant parameter and there is freedom in choosing the combination of operating wavelength and focal ratio of the optical system to match the desired Q factor consistent with the emulation goals. Raw images from space-based weak lensing missions will generally be undersampled and will require dithered exposures to reconstruct oversampled images (see section 3.3). For instance, the WFIRST DRM1 would have a focal ratio of f/15.9 and use Teledyne H2RG detectors with an 18µm pixel pitch; at $\lambda$=1.45µm (the blue edge of the DRM1 H-band), the images will have Q=1.28.[36] The WFIRST-AFTA design, whose 2.4m telescope would have a focal ratio of f/7.8 and use Teledyne H4RG detectors with a 10µm pixel pitch, has Q=1.08 at $\lambda$=1.38µm (the blue edge of the AFTA H-band). For the main results of this paper, our projector was set at f/21.9 and $\lambda$=0.88µm, which gives us Q=1.07.

In this paper, we only report on data obtained using a 50 mm x 50 mm chrome-on-glass target mask composed of four grids of 3 µm diameter holes. The mask also has a central cross pattern of spots for alignment purposes (see Figure 4). The grid spacing was chosen to be 650µm – not an integer multiple of the 18 µm pixel pitch – so that the spot centroid position (within a pixel) varied from spot to spot. We used a circular pupil stop in our shape measurement experiments that produces an f/21.9 beam. Due to machining tolerances, the pupil stop had a residual ellipticity of ~0.8%. At this focal ratio the sampling factor is Q=1.07 – within the weakly undersampled regime. By using point source targets instead of extended objects, the images in this initial emulation are more sensitive to aliasing, which is partially attenuated by smooth galaxy profiles. Point sources also provide a simple underlying ellipticity correlation function for us to measure (it should be zero).

---

[36] Other WFIRST designs will also have Q in the weakly undersampled regime.

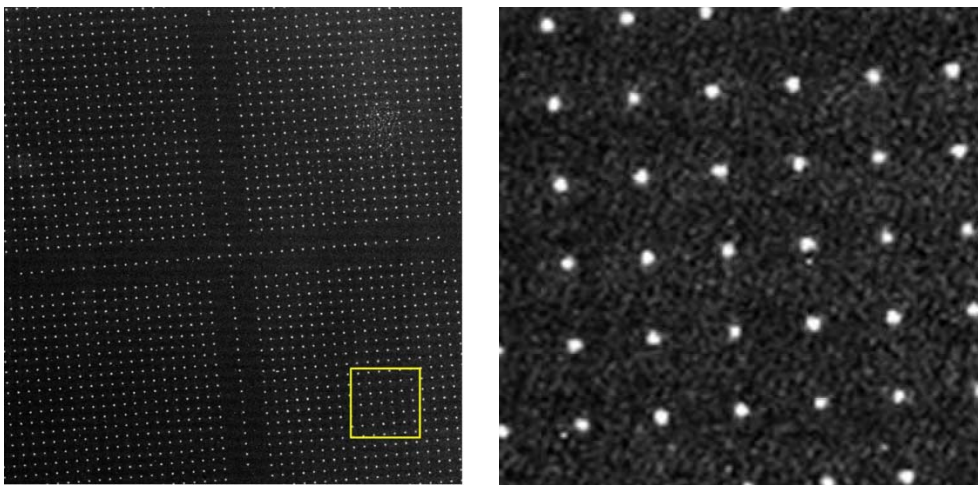

**Figure 4.** *(Left)* Image of a grid of 3μm pinhole spots projected onto the detector using a 0.88μm LED and an f/21.9 pupil stop. The spot grid spacing is 650 μm. Mask alignment is aided by the central cross and markers at its tips. *(Right)* Magnified view of the region outlined in yellow with grayscale adjusted to show the background level.

Typical WFIRST observations will consist of 150-250 second exposures (depending on the DRM), but a much shorter exposure time is used in this study to avoid unnecessary overheads. Lamp intensity can be changed to modify the overall signal-to-noise ratio (SNR) in our images to match WFIRST observations as and when necessary. Our 1 second exposure time yielded a typical SNR of over 200 (derived from the peak pixel value in a spot), indicating that our data is shot-noise limited. When emulating dithered exposures, the dither pattern was limited to mask translations within a 5x5 pixel box. We read the H2RG detector at a pixel rate of approximately 90kHz using 32 channels in correlated double sampling mode.

## 3. ANALYSIS PIPELINE

Figure 5 illustrates the customized data analysis pipeline that has been developed to allow rapid reduction of the data from raw images to various analysis products. The pipeline is divided into four stages: calibration, feature recognition, image reconstruction and analysis. The purpose of the first two stages is to create calibrated frames and a catalog of spot images that pass various quality criteria. The catalog is used together with the calibrated frames to reconstruct oversampled images of the spots. These reconstructed images are then passed on for further analysis, including determining the optimal focus position for each spot in a through-focus data cube, measuring the shapes of focused spots, and calculating correlation functions of the spot ellipticities. Each step is discussed in detail below.

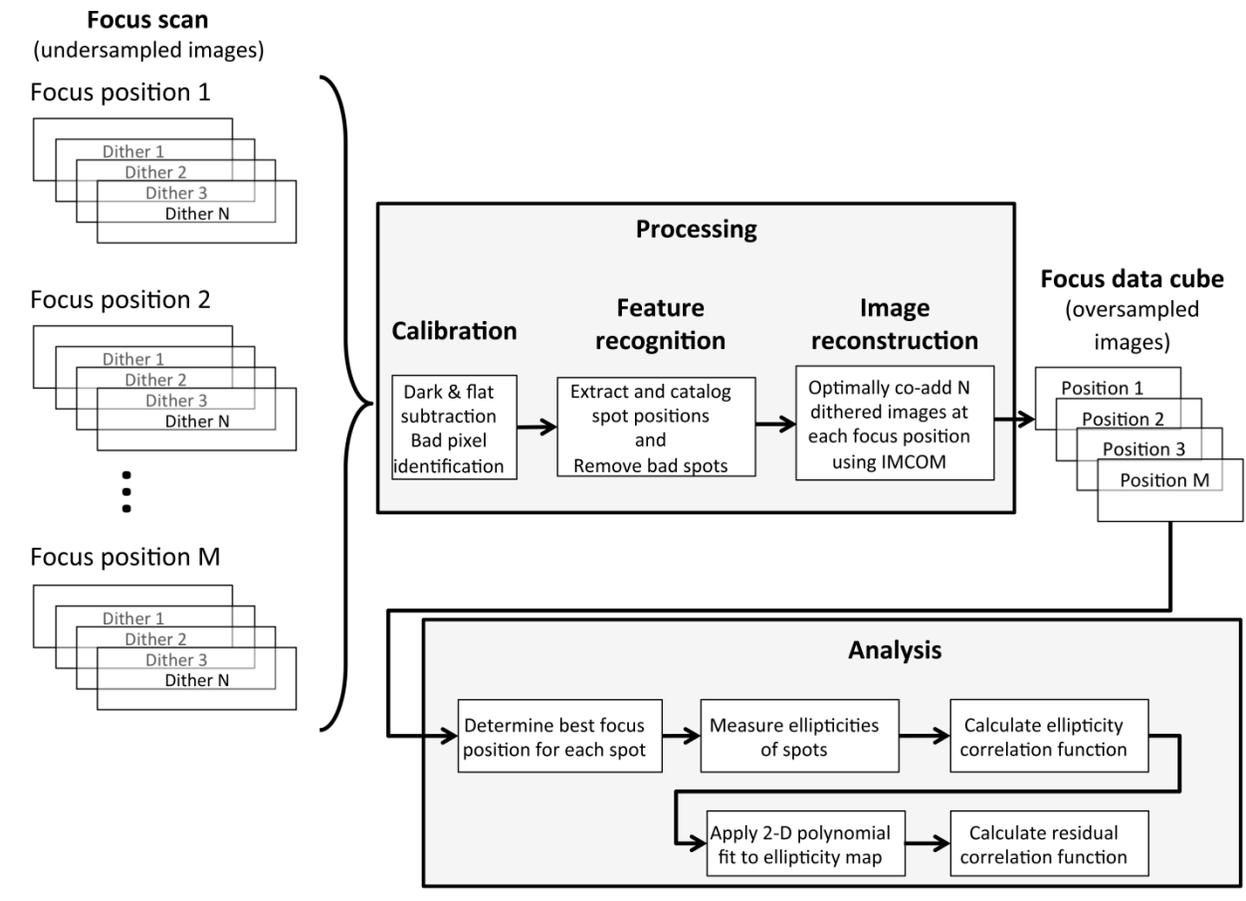

**Figure 5: Schematic diagram of the data analysis pipeline for the emulation experiments.**

Many of the pipeline steps are trivially parallelizable and are distributed over individual nodes of our computer cluster using the *dispy* Python framework.[37] This feature allows the reduction and analysis of large sequences of data on short timescales. The time-limiting step of the system is data acquisition, which varies depending on the exposure time, number of required dithers, and number of focus positions. A typical experiment in our current configuration takes several hours. It consists of stepping the mask through ~100 focus positions, taking 15 dithered exposures at each position, and analyzing ~1700 spots per exposure for a total of ~2.5 million spot images (about 50 GB of data) which must be processed. Consequently, our 12-node (144 CPU) computer cluster and 50 TB RAID storage are integral parts of our experimental setup.

---

[37] http://dispy.sourceforge.net/

## 3.1. *Calibration*

The raw images are initially processed using standard Image Reduction and Analysis Facility (IRAF)[38] routines, to perform dark subtraction, flat fielding correction and bad pixel identification. The required calibration data (bias, dark and flat field frames) are taken immediately prior to the experimental data. The individual bias, dark and flat-field calibration frames are generated by averaging ten frames of each type together using iterative sigma-clipping to remove outliers. Bad pixel masks are created by combining masks generated from three procedures: (i) identifying pixels above a 2-sigma threshold level in dark frames; (ii) non-linear behavior in the ratio of two flat field frames of different exposure times; (iii) interactive identification.

## 3.2. *Feature Recognition*

We use SExtractor[39] to find all spots within an exposure. The source catalogs are then further processed in two stages to remove misidentified spots. The first stage is to use SExtractor's own measurements to remove spots that fail our test criteria. The criteria include overall spot size, flux and illumination level and source classification. The limits on the first two criteria were set by trial and error to remove spurious detections such as bad/hot pixels and other detector artifacts such as scratches. The second stage is to remove any spots within a 5-pixel box around known bad pixels since the bad pixels will skew the ellipticity measurement for those spots. Such spots are removed from further consideration (rather than relying on interpolative correction of the bad pixel).

---

[38] http://iraf.noao.edu/
[39] Bertin, E. and Arnouts, S. 1996, ApJS, 117, 393

The image reconstruction and analysis procedures require a consistent catalog of spots associated with the entire experimental sequence. The final stage in feature recognition is therefore to "clean" the spot catalogs from individual frames and create a master catalog using an algorithm that identifies a common set of spots from frame to frame. The algorithm works by first defining a reference set of positions for all spot catalogs in a sequence of frames; typically this reference is simply the set of spot positions in the first frame of the sequence. For each subsequent "test" set of spot positions, a KDTree (using the *scipy*[40] python library implementation) is generated which allows an efficient nearest neighbor search to be performed between the test set and reference. The algorithm accounts for translation and rotation between the test set and the reference before calculating the nearest neighbor positions. The results are then filtered according to whether a nearest neighbor is within a certain (user-defined) distance of the expected position, producing for each frame a catalog of positions for only those spots identified in all frames. Under small translations and rotations, very few spots from the reference frame are lost; however, under large translations and rotations, a significant number of spots can be lost by moving out of the detector field of view. The tracking software is also capable of finding specific spot patterns (e.g. a cross of points in the mask).

---

[40] http://scipy.org

Both the image reconstruction procedure and the correlation function calculations rely on accurately measuring centroids for the target objects in the image. Spot centroids measured from noisy, undersampled frames will not be accurate to significantly better than a half pixel. However, individual spot positions are not needed for creating over-sampled images from the dithered, undersampled input. Rather, accurate knowledge of the dither pattern is required. The dithers consist only of mask translations; rotations and grid distortions are negligible. Relative translations are measured by calculating the change in spot centroids, averaged over all useable spots in a frame. Thus, errors in spot centroid determination in the raw images are reduced to negligible fractions of a pixel due to the √N effect of averaging over a large number of spots (N is typically several thousand).

### 3.3. *Image Reconstruction*

Accurate focusing of the target mask and shape measurements require reconstruction of oversampled images from the dithered, undersampled images. This step is an integral part of the pipeline. We use the IMage COMbination (IMCOM)[41] algorithm to optimally co-add multiple undersampled images to create an oversampled output image (see Figure 6). The algorithm produces the reconstructed images while minimizing both the final noise covariance of the pixel values and distortions to the image PSF. The undersampled source images used in the reconstruction need only differ by small displacements (i.e. dithers) in xy position on the detector. Our simulations show that the IMCOM algorithm itself has negligible effects on shape measurement and is only limited by one's ability to determine the relative dither displacements. [42]

---
[41] Rowe, B.T.P., Hirata, C. & Rhodes, J. D. 2011, ApJ, 741, 46

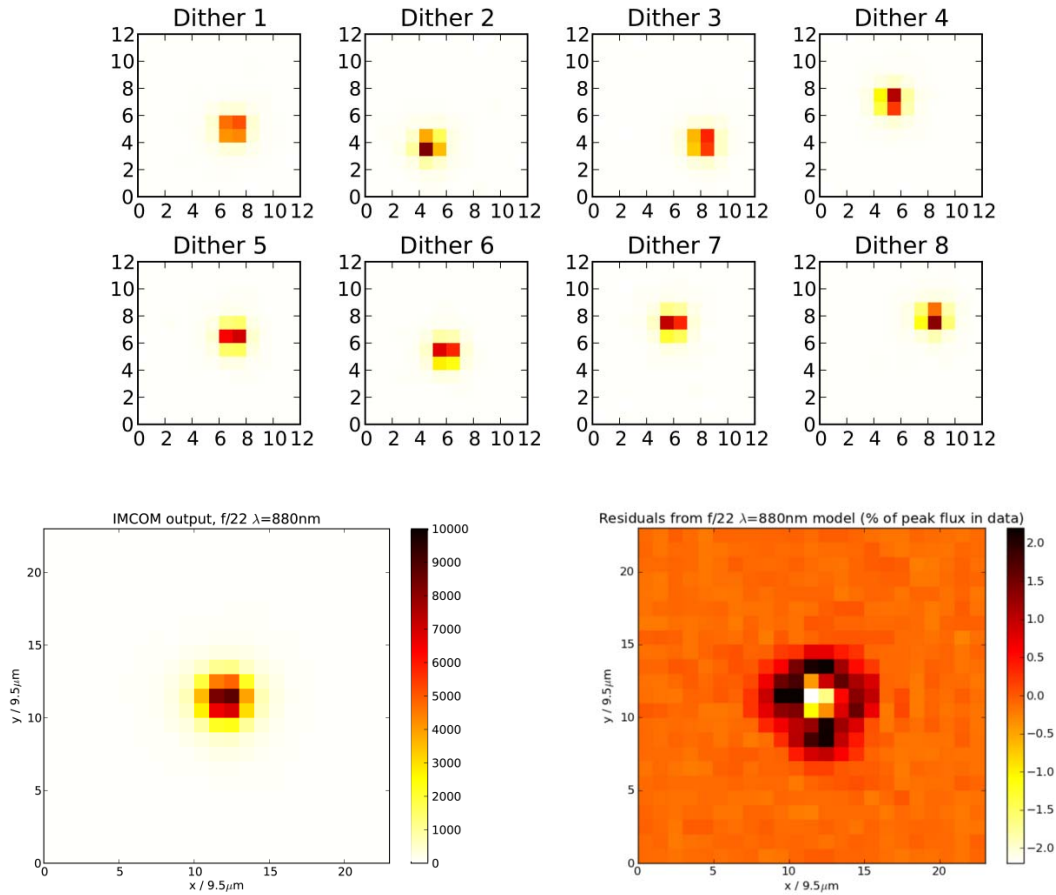

Figure 6: Demonstration of the image reconstruction process using the IMCOM algorithm. The intensity scales are in Digital Units. *Top eight panels:* Images of a single undersampled spot from 8 of 15 mask dithers projected onto the detector. The 20μm FWHM spots are sampled by 18μm pixels (Q=1.07). *Bottom panels: Left* – Over-sampled image reconstructed by IMCOM into 9.5μm pixels (Q=2.03). *Right* – Difference between the reconstructed image and a simplified PSF model that includes diffraction, pixelization and a Gaussian filter with σ=4.0μm. The Gaussian component accounts for blurring effects including charge diffusion (within the detector), residual defocus, and image motion. We have not partitioned these effects.

---

[42] C. Shapiro, B. T. P. Rowe, T. Goodsall, C. Hirata, J. Fucik, J. Rhodes, S. Seshadri, R. Smith, in prep.

The principal practical difference between IMCOM and the image combination approaches employed by Swarp[43] and Drizzle[44] is in the handling of undersampled data. Heavily undersampled imaging is not a regime for which image combination packages such as Swarp or MultiDrizzle[45] are designed, or in which they can operate without introducing distortions. These distortions arise due to the fact that both Swarp and Drizzle combine images via interpolation (with optional additional smoothing) across the input data: if these data are undersampled, the interpolated output will necessarily contain errors due to aliasing.[46] As demonstrated in Fruchter (2011)[47] aliasing defects can be large, even in weakly undersampled Hubble Space Telescope images, and such artifacts would prevent the characterization of detector-induced systematics at the level of precision required for dark energy science from missions such as WFIRST. Rather than interpolate across undersampled data, IMCOM instead solves for a general linear combination of input pixel values to produce oversampled output. The solution is found by extremizing a cost function that is specifically designed to control for any unwanted distortion in the final image, including aliasing. The generalized approach used in IMCOM can reconstruct images free from aliasing and has been demonstrated to do so at high precision.[42] It should be noted that the linear solution to the aliasing problem presented in Fruchter (2011) lies within the search space of the IMCOM algorithm.

---

[43] Bertin, E., Mellier, Y., Radovich, M. et. al., 2002, ASP Conf. Ser. 281, D.A. Bohlender, D. Durand, T.H. Handley eds., 228
[44] Fruchter, A.S. and Hook, R.N. 2002, PASP, 114, 792, 144
[45] Koekemoer. A.M., Fruchter, A.S., Hook, R., et al, 2003, Proc. of 2002 HST Calibration Workshop, STSci, Baltimore, Md, S. Arribas, A. Koekemoer and B. Whitmore eds., 337
[46] See, for example, Marks, R. J. 2009, *Handbook of Fourier Analysis & Its Applications*, Oxford Univ. Press.
[47] Fruchter, A.S. 2011, PASP, 123, 902, 497; arxiv: 1102.0292

IMCOM's ability to reconstruct oversampled images from undersampled data enables the study of intra-pixel response and sub-pixel PSF distortions – a driving motivation in our ongoing investigation. In addition, it permits accurate determination of centroid positions and grid distortion across the detector with sub-pixel resolution. IMCOM is also of great importance in accurately determining the optimal focus position of the optical system for undersampled PSFs. When using undersampled images, the concentration of a large fraction of total flux within a single pixel renders most metrics of image quality (and thus focus) useless. We discuss the methodology for determining the focus in section 3.4.

To function, IMCOM requires accurate knowledge of the relative displacements between the source image frames used to reconstruct the final output. This is derived from the consistent catalogues of point sources derived from sequences of dithered images. IMCOM provides a check on whether the input dither images were sufficient to realize a oversampled output (in the form of the output $U_\alpha$ and $\Sigma_{\alpha\alpha}$ as defined in Rowe, et. al. (2011)[41]). In addition, IMCOM requires an input model PSF (generated from nominal system parameters) and a set of tolerances defining the final quality of the reconstruction. Our pipeline outputs a data cube of spot images reconstructed by IMCOM. Each image is a small "postage stamp" with the reconstructed spot near the center (see lower left panel of Figure 6). The size of the postage stamp varies depending on the pixel scale in the reconstruction. For the main results of this paper, we use 24x24 pixel postage stamps with 9.5µm pixels for oversampled images of spots with a typical FWHM of 20µm.

The lower left panel of Figure 6 shows an example of the reconstructed output of a single point source generated from multiple dithered input images. The original images are sampled with the detector's 18µm pixels (Q=1.07) and converted into an oversampled 9.5µm pixel image (Q=2.03). We compare the output to a noise-free toy-model of our system that includes only diffraction, charge diffusion, and an ideal pixel response. The small residual (shown in the lower right panel of Figure 6) of ~0.5% demonstrates that this simplified model accounts for most of the PSF. This model is not used in our analysis and is only shown here for illustrative purposes.

### 3.4. *Focus*

A well-focused image is crucial for accurate shape measurements because the ellipticity of a target object in the image plane can be strongly dependent on the amount of defocus, especially in the presence of other optical aberrations - particularly astigmatism.[48] We show how our ellipticity measurements vary with defocus in section 4.3.

Focus is achieved by first obtaining images of the target mask at several positions along the optical axis (a focus scan). The optimal focus position is then identified for each spot in the target mask from this data set, using a suitable metric (see below). In practice, all spots are not in focus simultaneously, so the set of best focus positions lies on an irregular 3D surface. For coarse focusing we use a plane fitted to this surface to adjust the mask position, bringing the mask and detector image planes into closer parallel alignment. This process is repeated until no further improvement is useful.

---

[48] Jarvis, M., Schechter, P., and Jain, B.; arXiv:0810.0027

We evaluated several different measures of the optimal focal surface. In each case the best focus position was determined by finding the extremum of a parabolic fit to some metric versus focus position, as expected for a near diffraction-limited system.[49] Figure 7 shows a comparison of 4 different metrics: the standard deviation of the pixel values in an aperture around each spot (panel A), the peak pixel value of each spot (panel B), the intensity weighted radius squared, $R^2$, of the spot (panel C) and the modulus of the ellipticity estimator $\sqrt{(\varepsilon_1^2+\varepsilon_2^2)}$ (defined in section 3.5) (panel D). To normalize the data to a common scale, each fit has been divided by the inferred minimum or maximum. Five representative spots across the mask are shown, with an artificial vertical shift applied to each spot for visual clarity. The standard deviation metric (A) is well fit by a 2$^{nd}$ order polynomial and has the lowest residual fit errors compared to any other metric. The ellipticity (D) is a poor focus metric since, in addition to being noisy, minimum ellipticity doesn't necessarily correspond to maximum focus. Defocus effects on the ellipticity pseudo-vector can be positive or negative and can counteract other sources of distortion such as pupil shape, source shape, or detector effects. Moreover, minimizing ellipticity for each spot may obscure the very distortions from the detector that we want to measure. By contrast, the other metrics are proxies for the spot profile, which can only increase with defocus.

---

[49] Ross, T.S. 2009, Applied Optics, 48, 10, 1812

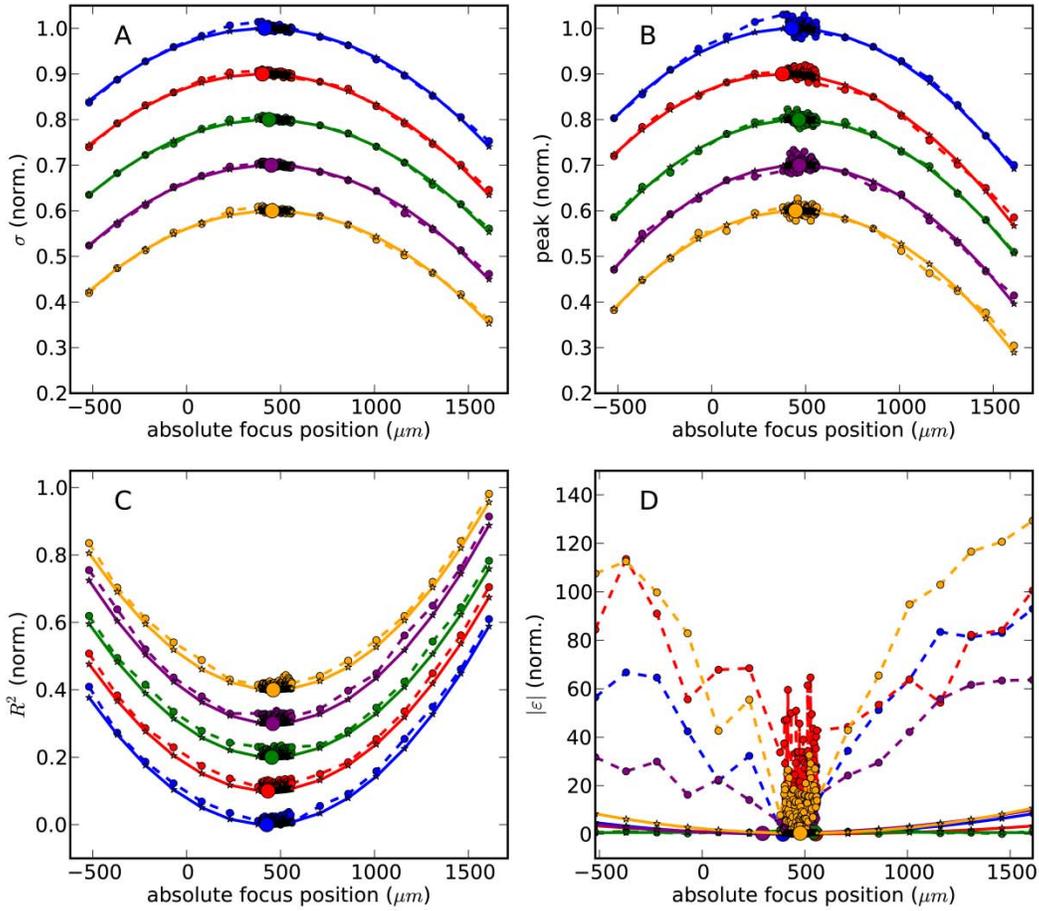

**Figure 7: Comparison of focus metrics from a representative focus scan for five different spots (corresponding to the offset lines of different color/shade). The dashed lines and circles represent the input data and the solid lines and stars the fit to the data. The offsets between the five different spots are artificial for visual clarity. The large circles in each curve show the best focus position for that spot. Each focus scan combines coarse 150μm steps with fine 2μm steps in the range where the best focus is expected. The four plots show focusing using four different metrics: the standard deviation of the pixel values in an aperture around each spot, the peak pixel value in a spot image, the intensity weighted radius squared ($R^2$) and the absolute ellipticity value $\sqrt{(\varepsilon_1^2+\varepsilon_2^2)}$. The standard deviation is reasonably well fit by a parabola, while the peak pixel value and $R^2$ metrics are slightly noisier. Ellipticity is clearly a poor focus metric.**

We adopted the standard deviation as our focusing metric because it is physically motivated and has lower residual fitting errors compared to more conventional metrics. It can be shown that the standard deviation of the pixel values is inversely proportional to the squared width, $\sigma^2$, of a 2D-Gaussian fit to the spot profile. Thus, as the spot moves closer to focus, and the width of the spot profile decreases, the standard deviation will reach a maximum. The increased fitting accuracy of the standard deviation metric over the Strehl-ratio (peak intensity) metric can be understood by considering what happens when a (reconstructed) spot is located near a pixel boundary. In this case the maximum pixel value is a poor estimate of the true peak flux as the photo-generated charge is distributed amongst neighboring pixels. When the pixel scale is comparable to the PSF size, as it is here, an accurate peak value can only be determined by fitting the flux distribution. This fitting requirement adds a large computational burden (for the thousands of point sources in the field) for this focusing technique. By contrast, the standard deviation metric is relatively inexpensive computationally. As further validation of this approach, we calculated the Strehl-ratio for the optimal focal surface using our adopted metric and found that it matched the expected value from optical modeling and from actual PSF measurements.

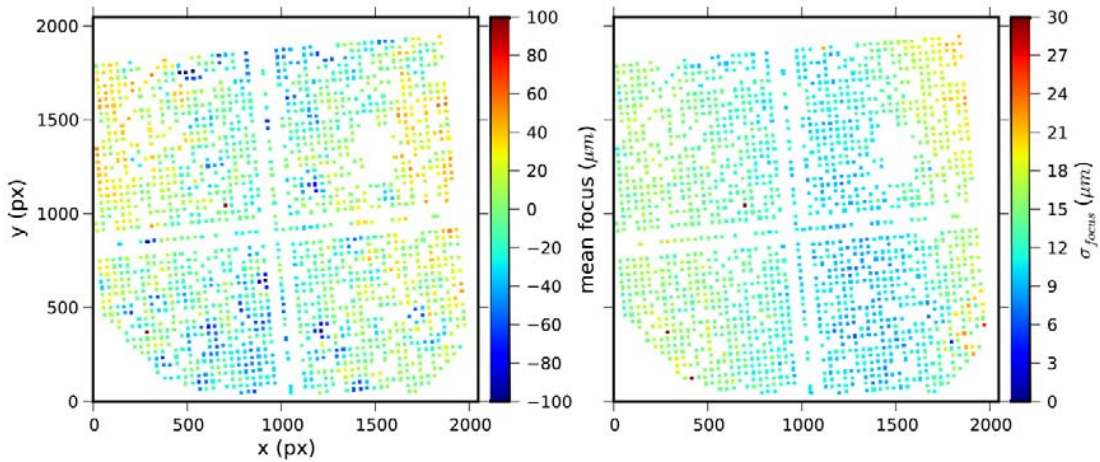

Figure 8: (*Left*) Mean and *(Right) s*tandard deviation maps of the best focus position of each spot, relative to motion stage focus position, obtained from five repeated focus sequences. The intensities correspond to relative differences in focus position across the detector. Data was obtained from a pinhole grid mask imaged using an f/22 pupil. The complex shape of this optimal focal surface cannot be adequately fitted with a plane. This drives the need to acquire data while scanning through focus.

The left panel of Figure 8 shows the optimal focal surface determined from the average of the five data sets described in this paper. The optimal focal surface of each data set was used when analyzing the data sets individually. There is a substantial amount of structure across this focal surface with a peak-to-valley variation of approximately ±50 μm that cannot be adequately described by a plane.  The source of this structure can be attributed to a number of different factors, including the surface flatness of the detector and mask, residual mask-detector non-planarity, other optical aberrations, and thermal variations (we do not partition these contributions).  We significantly reduce focus errors by retaining all spot images in a fine focus scan (2μm steps) and then analyzing individual spots (e.g. measuring ellipticity) from selected images that were generated nearest to each spot's best focus position along the optical axis. Although these acquisition steps lead to a much larger data volume (factor of 100), they are necessary to avoid the spurious ellipticity that results from analyzing the data from a single optimized focal plane (see section 4.3).

## 3.5. *Ellipticity Measurement*

Astronomers use image ellipticity as an estimator for shear. Note that these are not synonymous concepts: ellipticity is a geometric property describing the light profile, while shear refers to a linear transformation applied to a galaxy image (see Figure 1). Ellipticity can be calculated using different conventions.[18] For the sake of conceptual and computational simplicity, we compute ellipticity using the second moments of the pixel values. If $I(\vec{r})$ is the intensity profile of the object, then our weighted ellipticity estimators, $\varepsilon_1$ and $\varepsilon_2$, are defined as:

$$Q_{ij} \equiv \frac{\int d^2 r I(\vec{r}) w(\vec{r})(r_i - \bar{r}_i)(r_j - \bar{r}_j)}{\int d^2 r I(\vec{r}) w(\vec{r})} \qquad \text{Equation 6}$$

$$\varepsilon_1 = \frac{Q_{xx} - Q_{yy}}{Q_{xx} + Q_{yy}}; \qquad \varepsilon_2 = \frac{2 Q_{xy}}{Q_{xx} + Q_{yy}} \qquad \text{Equation 7}$$

where i, j correspond to either axis of the pixelated image, and $\bar{r}_i$ is the weighted image centroid (1st moment). The Gaussian weighting function $w(\vec{r})$ is introduced to ensure that the integrals converge in the presence of noise. The center of the function must coincide with the weighted centroid – this is done iteratively. We compute the integrals by simply summing over pixel positions without any interpolation, finding that any numerical error so introduced is negligible compared to shot noise in the images. The sum is taken over the postage stamp image output by IMCOM. Before summing, we subtract off a constant background, which is estimated by taking the mean pixel value in a 3-pixel border around the postage stamp.

Error tolerances for weak lensing surveys are quoted in terms of gravitational shear, not ellipticity. Therefore, we need a calibration to relate our ellipticity estimator – which includes the effects of the optics, detector, and the weighting function – to the effect of gravitational shear alone. A realistic calibration would be a function of the source properties such as shape and signal-to-noise – this is called *shear susceptibility*.[50] For our purposes, a crude scaling factor, $m = \gamma_i / \varepsilon_i$, is sufficient to relate our measurements to weak lensing shear requirements.

---

[50] See, for instance, Leauthaud, A., Massey, R., Kneib, J-P., et al., 2007, ApJS, 172, 1, 219

In our emulation, point sources are useful since any non-zero ellipticity is a measurement error. However, point sources don't experience gravitational shear. Therefore, to put the measurement in context, we use a calibration to approximate the shear bias that such an ellipticity error would induce in a galaxy image, making the conservative assumption that detector-induced distortions will affect point sources more strongly than resolved galaxies. Moreover, in a real weak lensing analysis, ellipticity measurements of point sources are used to correct the shears of nearby galaxies. So our calibration can also be interpreted as an approximation of the spurious shear correction that would be induced by an ellipticity measurement error.

We compute the calibration factor by simulating images with known shears and measuring the ellipticity of those images after charge diffusion, pixel response, and the weight function are included. For our definition, with the weighting function, $w(\vec{r})$, set to 1, a radially symmetric image sheared by $\gamma_i$ will have an ellipticity $\varepsilon_i = 2*\gamma_i$.[51] Our radially-symmetric, Gaussian weighting function will reduce the ellipticity of asymmetric images. Charge diffusion in the detector and pixelization also reduce the measured ellipticity of an elliptical source. The latter does this by averaging the light profile over finite pixel areas. We simulated images of our projector's PSF (operating at f/22 and $\lambda$=0.88 µm) and made them elliptical (up to $\varepsilon$ < 0.3) by rescaling the image coordinates and including the above effects. We find that they reduce ellipticity by an approximately constant factor of 1.85. Thus, our final calibration is

$$\gamma_i = \frac{1.85}{2} \varepsilon_i \qquad \text{Equation 8}$$

---

[51] Dodelson, S. 2003, *Modern Cosmology*, Academic Press

or m = 0.93. Variations in the charge diffusion and pixel response across the field of view may cause distortions that will propagate into our final measurement of the shear correlation function. However, we are unconcerned with any effects that can be accounted for by this uniform, linear calibration curve.

## 3.6. *Shear Correlation Functions*

In order to relate our measurements to weak lensing observables, we compute spatial correlation functions for the ellipticity maps. Our correlation function is simply the mean product of the corresponding shears of all spot pairs at a given separation $r = |\vec{r}|$:

$$\xi_{ii}(r) \equiv \frac{1}{N_r}\Sigma_{mn}\,\gamma_i(\vec{r}_m)\gamma_i(\vec{r}_n) \qquad \text{Equation 9}$$

where i denotes the shear component ($\gamma_1$ or $\gamma_2$), and $\vec{r}_m$ and $\vec{r}_n$ are the spots' xy-coordinates on the mask. The sum is over all pairs of spots (m,n) with a separation $r = |\vec{r}_m - \vec{r}_n|$ within a bin around r. $N_r$ is the total number of such pairs in the bin. We use evenly spaced logarithmic bins ($d\log_{10} r/\text{pixels} = 0.132$) with a minimum separation of r=30 pixels on the lowest bin, just below the 36.11 pixel spacing of the spot grid. The above definition is a monopole correlation function, which does not depend on the orientation of the spot pairs. This computation is analogous to the shear correlation that astronomers measure for galaxies or stars as a function of angular separation on the sky.[52]

---

[52] See e.g. Massey, R., Rhodes, J., Leauthaud, A., et al., 2007, ApJS. 172, 1, 239

Our definitions of $\gamma_1$ and $\gamma_2$ are relative to our detector orientation; however, it is convenient to define a more symmetric basis that is independent of detector coordinates and corresponds, instead, to the relative coordinates specified by the spot separation vector, $\vec{s} = \vec{r}_m - \vec{r}_n$. Translating the measured ellipticities into this basis yields tangential and cross components of the ellipticity, relative to this separation vector. If $\psi$ is the angle between $\vec{s}$ and the x-axis of the detector, then

$$\gamma_t \equiv -\gamma_1 \cos(2\psi) - \gamma_2 \sin(2\psi) \quad \text{Equation 10}$$

$$\gamma_\times \equiv -\gamma_1 \sin(2\psi) - \gamma_2 \cos(2\psi) \quad \text{Equation 11}$$

We can then define new, detector coordinate-independent, correlation functions:

$$\xi_\pm(r) \equiv \frac{1}{N_r} \sum_{mn} [\gamma_t(\vec{r}_m)\gamma_t(\vec{r}_n) \pm \gamma_\times(\vec{r}_m)\gamma_\times(\vec{r}_n)] \quad \text{Equation 12}$$

where $\gamma_t$ and $\gamma_\times$ must be calculated for each spot pair. In real weak lensing analyses these functions are useful since they are more directly related to the physics of the gravitational signal and can aid in troubleshooting systematics.

# 4. Demonstration of imaging capability

## 4.1. *Quality of Image Reconstruction Across Sampling Regimes*

To verify the performance of our optical system, we measured the sizes of reconstructed spots and compared them to a simple PSF model. Accounting for diffraction alone, the full-width at half-maximum (FWHM) of the PSF is

$$\text{FWHM} \approx 1.03 * \lambda * f \qquad \text{Equation 13}$$

where $\lambda$ is the wavelength of the illumination source and $f$ is the focal ratio. We expect this relation to hold at large $\lambda$ and/or focal ratio, where the PSF is dominated by diffraction. For smaller PSFs, we expect deviations due to pixelization, interpixel crosstalk, and charge diffusion in the detector, all of which increase the PSF size. We modeled the PSF as a convolution of a diffraction-limited Airy function, an 18µm square boxed pixel response function, and a term to account for diffusion of photo-generated charge. We approximate the charge diffusion term, which is described by a hyperbolic-secant function, by a Gaussian with σ=2.94µm; this width corresponds to a diffusion length of 1.87µm measured previously on the same H2RG device.[29]

Figure 9 demonstrates that the system is working as designed with a PSF that is reasonably approximated by our simple model. Dithered images of the spot grid were taken using 3 pupil stops, corresponding to f/11, f/21.9 and f/44. SExtractor was then used to measure FWHM for all useable, reconstructed spots in a single focal plane (1s exposures). For each focal ratio, we plot the mean and variance of the measured FWHM. The variance in the f/44 data is dominated by shot noise, while the f/11 data is more sensitive to focus variations (deviation of the focal surface from a plane). The three measurements are compared to the dashed line, which we computed by running SExtractor on simulated point source images generated from our simple model with very high signal-to-noise. For reference, we also plot Equation 13, which holds in the diffraction limit; however, we expect a discrepancy since SExtractor assumes a Gaussian profile, whereas the FWHM in Equation 13 refers to that of an Airy function.

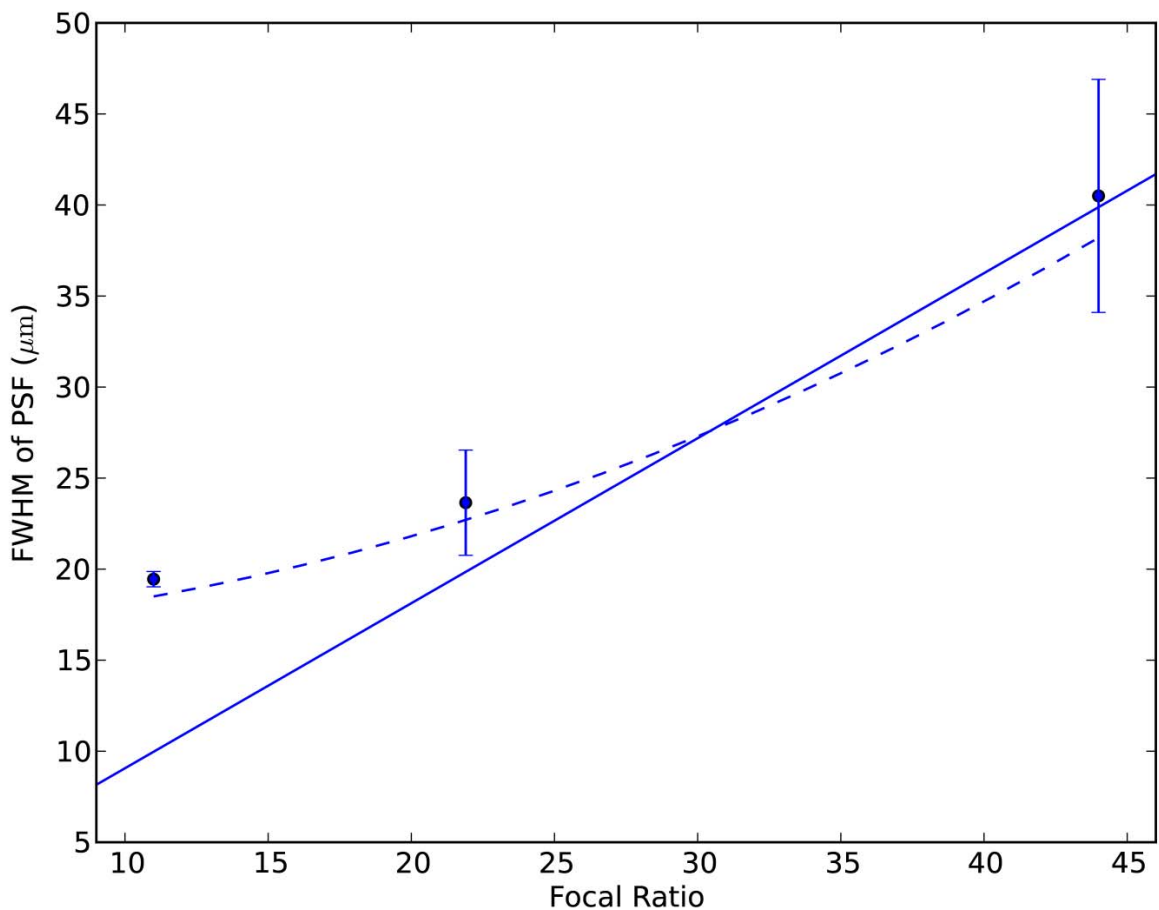

Figure 9: FWHM of the PSF as a function of focal ratio for $\lambda$ = 0.88µm. The solid line is the expected FWHM of an Airy spot (diffraction only). The dashed line is generated by SExtractor measurements of PSFs simulated using a simple model that includes diffraction, pixelization and charge diffusion. Note that SExtractor underestimates the width of an Airy function since it assumes a Gaussian profile. The data points and error bars are the means and standard deviations of SExtractor measurements of reconstructed images in a single focal plane (1s exposures).

## *4.2 Focus Variability and Reproducibility*

The right panel of Figure 8 illustrates the variability in the optimal focal surface (see section 3.4) across data sets. The source of this variability is presently unknown, but we suspect it is related to thermal instability. It manifests not just as a constant offset (i.e. a change in optical path length), but it also has structure across the field. Our focus method insulates us from this variability since we re-measure the optimal focal surface for each new data set and select spot images generated close to that surface. Nevertheless, any residual defocus will create a coherent ellipticity pattern and bias the measured ellipticity correlation functions. Since the main result of this paper is an *upper limit* on the amplitude of detector-induced correlation functions, our result is robust to this type of error.

## *4.3 Variation of Ellipticity with Defocus*

Systematic ellipticity signatures can be induced in the data by optical misalignments (among other factors). In particular, the amount of ellipticity is proportional to the product of defocus and astigmatism.[48, 53] In Figure 10, we plot ellipticity versus defocus, illustrating that our data are consistent with the expected trend. The ellipticities have been averaged over all spots with the same defocus in five focus scans. For an ideal optical system with a circular pupil, the individual ellipticity components should change sign as the point source images pass through focus. The non-vanishing ellipticity seen at zero defocus is at least partially caused by machining error on our pupil (see section 5.1). The larger variations in $\varepsilon_1$ than $\varepsilon_2$ can be explained by symmetries of the Offner optics. To first order, ellipticity is proportional to the product of defocus and astigmatism. When well-aligned, aberrations in the Offner optics are dominated by a small astigmatism which elongates the image either in the radial or tangential direction. In our present projector configuration of the detector, this corresponds to the row and column direction and is thus purely $\varepsilon_1$ along the centerline. At other field points, the transformation from the radial/tangential elongation to detector coordinates produces a very slight $\varepsilon_2$ component so that the mean $\varepsilon_2$ over a full image is no longer zero. We expect the observed variation of $\varepsilon_1$ and $\varepsilon_2$ with defocus to change as the detector is rotated, relative to the optics, but have not yet performed those experiments.

---

[53] Noecker, C., 2010, Proc. SPIE, 7731, 77311E

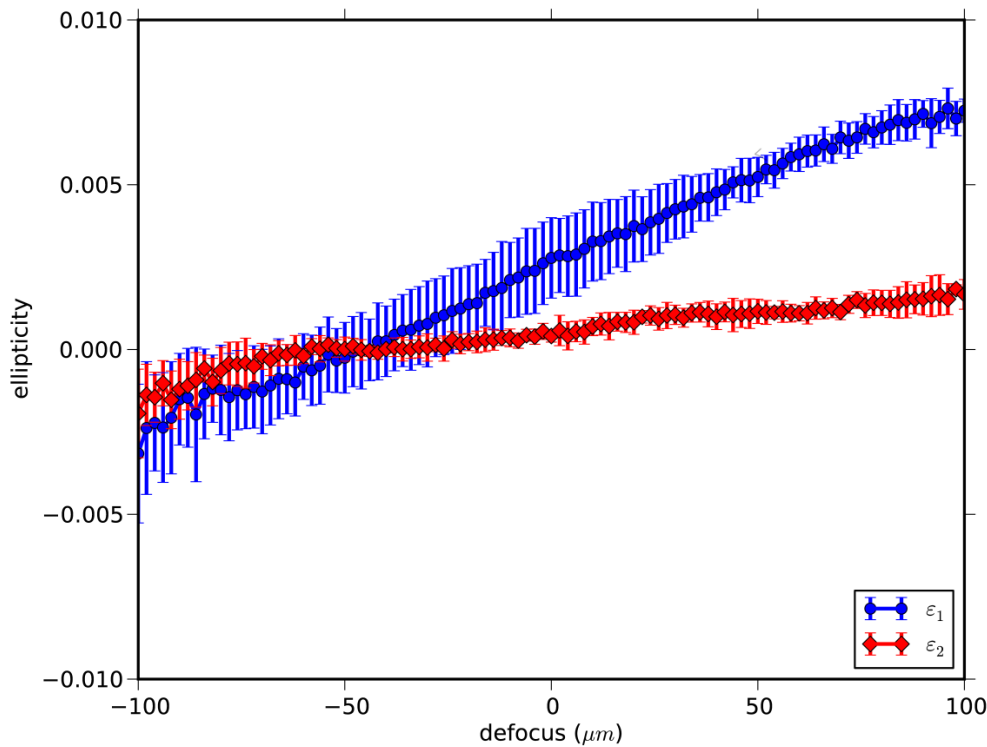

Figure 10: Uncalibrated measurements of mean ellipticity versus defocus. The error bars show the standard deviations over 5 focus scans; they are highly correlated since the trend varied from scan to scan. For circular sources and a circular pupil, we expect $\varepsilon_1$ and $\varepsilon_2$ to be proportional to defocus. The differing variation of the two ellipticity components with defocus is hypothesized to be due to asymmetry in the astigmatism of the Offner optics (see text).

## 4.4. *Image Motion*

Shape measurements may be sensitive to image motion artifacts due to refractive effects (referred to as astronomical "seeing") along the line of sight, mechanical vibrations, thermal expansion and contraction of system components, etc.  Effects that have a preferential direction with respect to the image may also induce an ellipticity.  We measured the system's sensitivity to such artifacts by measuring the mean relative displacement of the spot grid in a rapid sequence of exposures.  The experiment consisted of reading a small sub-region (a band encompassing a single row of spots in the mask) of the detector for two minutes in a high frame-rate readout mode (10ms per exposure) with a cadence of 24ms between exposures, much shorter than our typical 1s exposure time. The frames were then passed through the analysis pipeline, and the average spot displacement from each exposure was calculated.

Figure 11 shows the temporal variability in spot position over two minutes. The overall motion is roughly axi-symmetric with deviations at the level of 500nm RMS, which is approximately 3% of a pixel.  By comparison, charge diffusion in the H2RG detector approximately convolves the PSF with a 2D Gaussian with width $\sigma$ = 2.94µm or 16% of a pixel.  PSF blurring due to image motion and charge diffusion will add in quadrature, making the former a relatively small effect. We verified this by co-adding simulated spots with relative positions given by sub-sequences of the high-frame rate data; the number of co-adds was set by the ratio of our typical exposure time to the high-speed cadence.

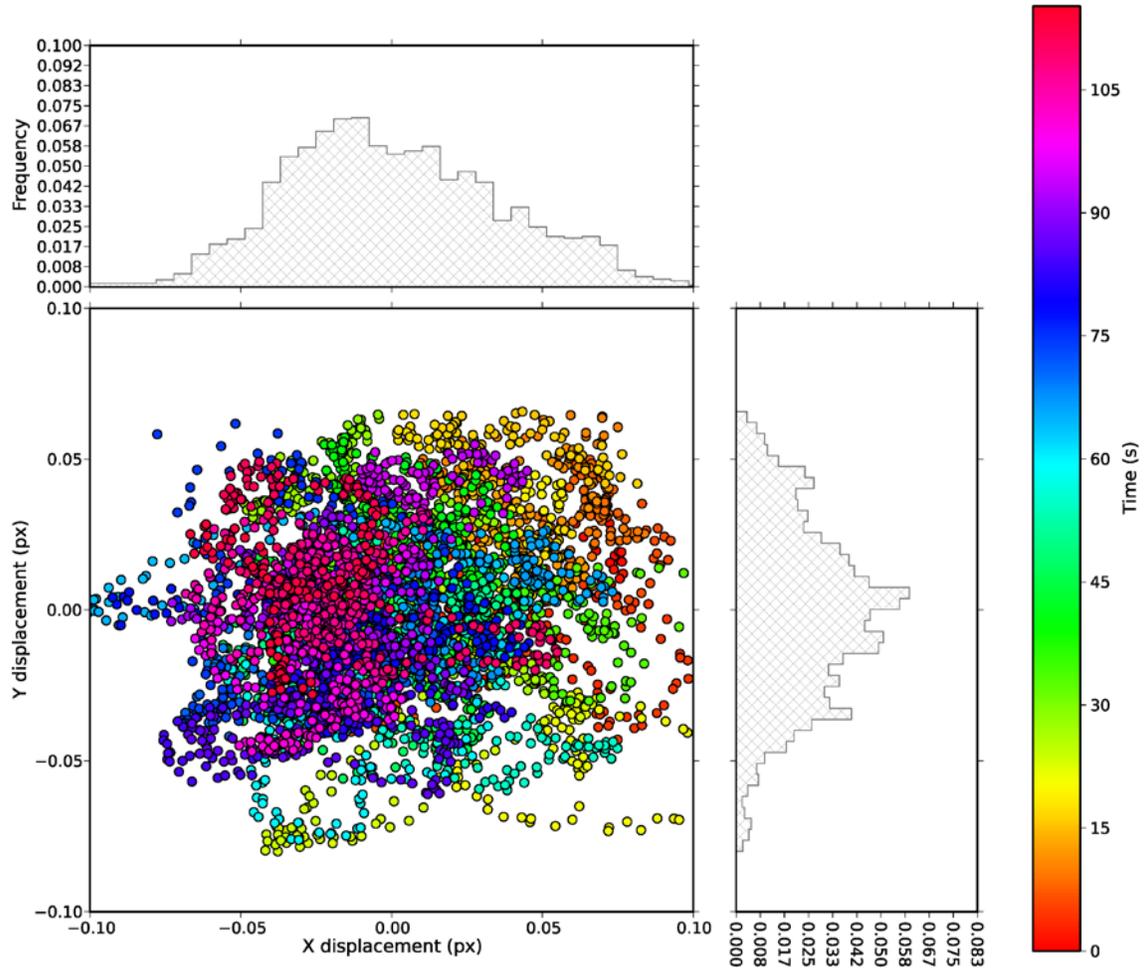

Figure 11: Mean centroid motion of all spots as a fraction of the pixel pitch. Each set of colored dots illustrates the range of displacements over 1-second time intervals. Histograms at the top and right of the figure illustrate the frequency of the relative displacement over the entire 2 minute exposure sequence.

Although image motion is largely isotropic over long timescales, non-random motion within a 1s exposure may slightly shear the PSF. Since we do not detect any relative motion between spot pairs over time, this shearing would be identical for all spots in a frame and could create correlated ellipticities. Fortunately, this effect will be partially averaged out by the image reconstruction process, which combines multiple dithered exposures. Since this paper reports an upper limit on shape distortions from the detector, our result is robust to this type of error.

Nevertheless, our future plans include system upgrades to reduce image motion, which will improve our ability to measure sub-pixel detector features.

## 5. Results

### 5.1. *Ellipticity Measurement*

A map of the measured, uncalibrated, ellipticity of each reconstructed spot from its best focus position is shown in the whisker plot in Figure 12. The length of each line represents the magnitude of the ellipticity of the object at that location. The orientation of the line represents the orientation of the ellipticity pseudo-vector. Bad cosmetics on the engineering grade detector are responsible for the data drop outs in the figure.  Note that because exposures are dithered in a 5x5 pixel box, coordinates on this plot do not correspond to unique positions on the detector – they should instead be interpreted as coordinates on the mask (or the emulated "sky").  Histograms of the ellipticity values from five repeated experiments are shown below. The ellipticities measured in a single scan are characteristically $O(10^{-2})$ due to random noise, particularly photon shot noise.  Although the mean squared ellipticity is thus $O(10^{-4})$, we will show that the correlation function for spot pairs is much smaller.

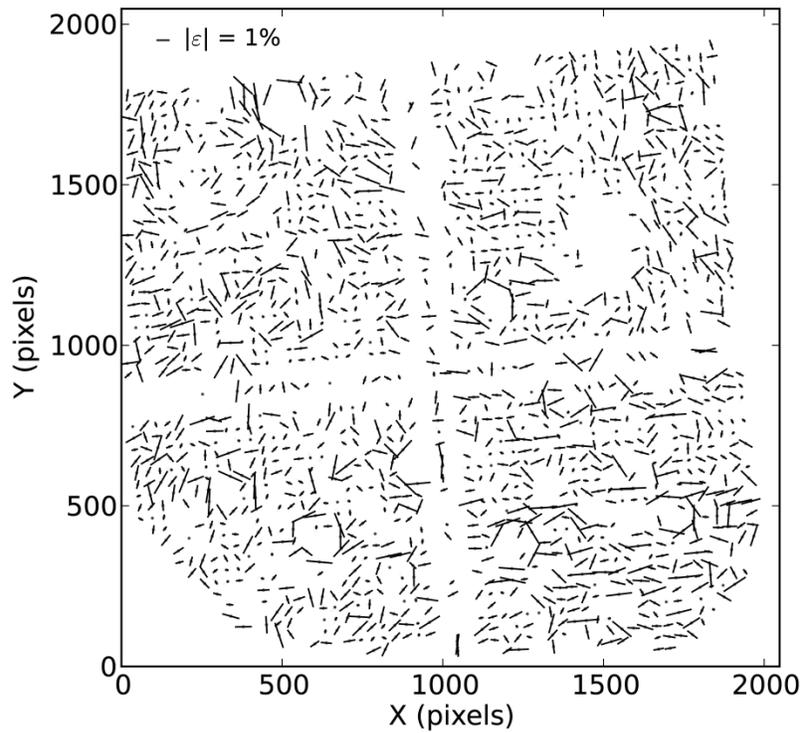

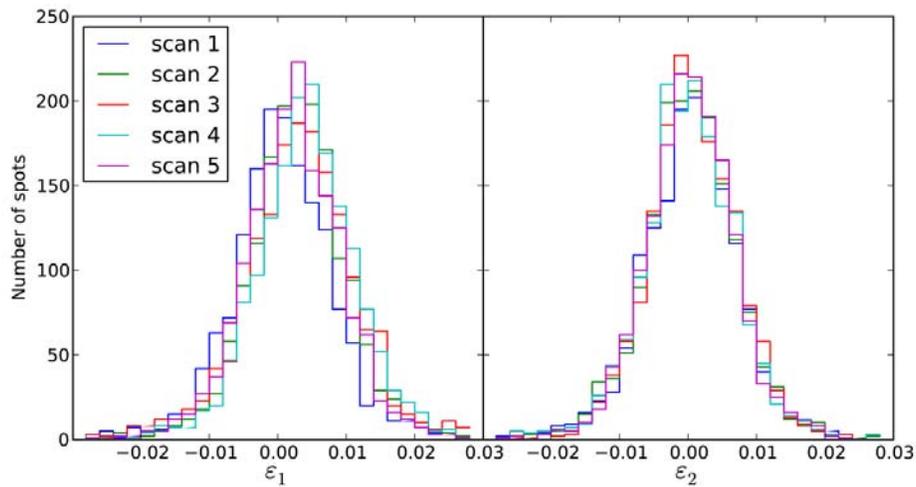

**Figure 12:** *(Top)* Whisker map for focus scan #1 depicting the magnitude and angle of the ellipticity for uncalibrated ellipticities of individual spots at the surface of best focus. Spots near bad pixels have been removed. For visual clarity, the largest size given to any whisker on the plot is 2%, and whiskers too small to see have been replaced by dots. *(Bottom)* Histograms of the ellipticity from five repeated scans. All scans have 1664 spots after cuts near bad pixels.

Table 2: Means and standard errors for uncalibrated spot ellipticities at the surface of best focus for five repeated measurements. There are 1664 spots in each scan after cuts for bad pixels. The full distributions are shown in Figure 12.

| Data set # | 1 | 2 | 3 | 4 | 5 |
| --- | --- | --- | --- | --- | --- |
| $\varepsilon_1$ (%) | .091±0.02 | 0.32±0.02 | 0.35±0.03 | 0.42±0.02 | 0.21±0.02 |
| $\varepsilon_2$ (%) | .096±0.04 | 0.064±0.03 | 0.11±0.04 | 0.037±0.03 | 0.045±0.02 |

Table 2 shows that the mean ellipticity is generally inconsistent with zero. The dominant source of the non-zero mean is likely to be the machining error of the f/22 pupil and residual mis-alignment of the secondary mirror relative to the primary. The pupil's semi-major and semi-minor axes correspond to a 0.8% ellipticity. Assuming the pupil is perfectly elliptical and aligned, it should produce spots with uncalibrated ellipticities of 0.43%. Simulations show that we are insensitive to shape errors in the 3 μm spot mask itself, with mask ellipticities as high as the 3% fabrication tolerance resulting in less than 0.1% uncalibrated ellipticity in the final image. This insensitivity occurs because the imaged pinhole is a convolution of the original target shape with the much larger PSF (20μm FWHM). Note that the effect of a constant ellipticity bias is easily removed from the correlation function.

## 5.2. Correlation Functions

Figure 13 shows the shear correlation functions for our five independent data sets. Figure 14 shows the distribution of spot pairs for each range of spot separations. Measurements using the smallest and largest separation bins are noisy due to a low number of pair combinations. Error bars in Figure 13 are estimated by computing the standard error of ellipticity pair products in each bin, which does not fully take into account correlations between the bins.

Some run-to-run variability is apparent in the correlation functions. We will show below that this variability is almost entirely due to large-scale patterns in the PSF. Assuming that the measured structure is completely due to the detector places an upper limit on the size of the detector-induced correlation functions.

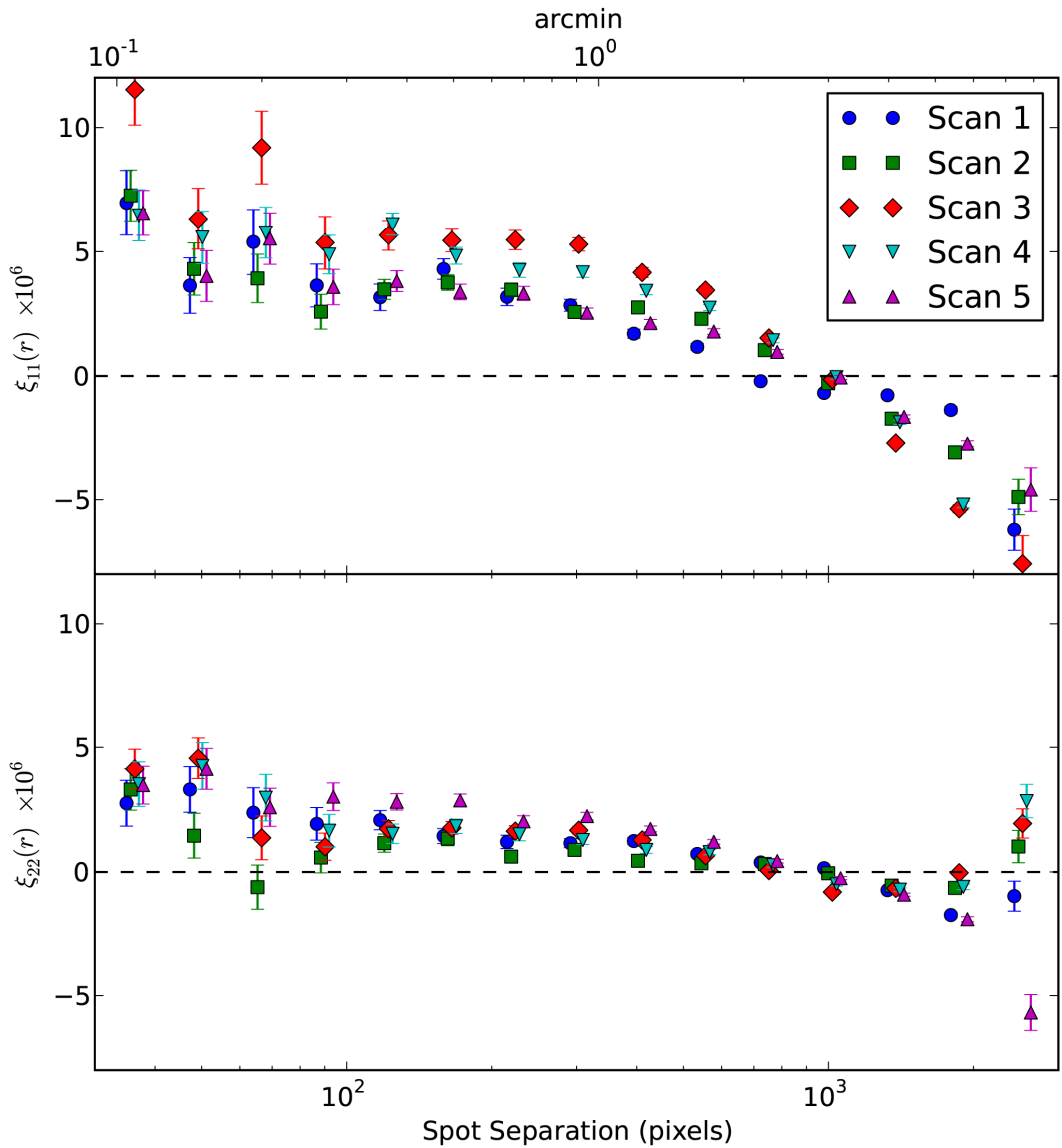

Figure 13: Plot of the correlation function versus spot separation distance *r* for five repeated measurements. Correlations are calculated in mask coordinates, using Eqn. 9, but we plot separations in units of detector pixels (18μm/pixel). The trends are predominantly due to large-scale patterns in the PSF (see text and Figure 15). The upper axis along the abscissa has been scaled to 0.18 arcminutes/pixel, which is the proposed scale for WFIRST DRM1.

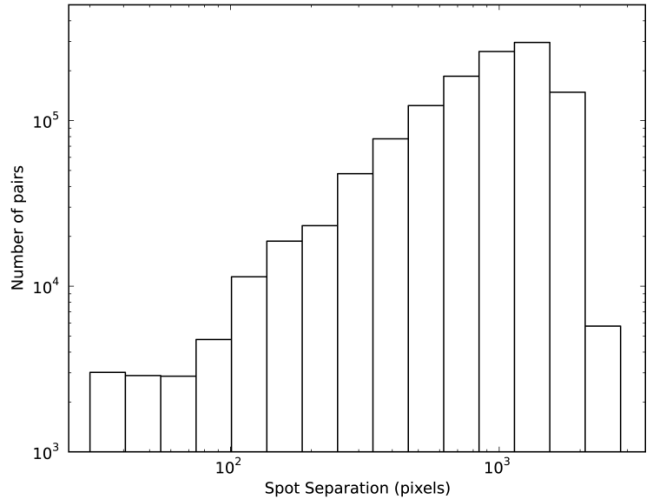

**Figure 14: Histogram of the number of spot pairs at each bin of spot separations in Figure 13.**

In computing the correlation functions, we eliminated spots with an $\varepsilon_1$ or $\varepsilon_2$ that are outliers by more than 4 standard deviations; at most 20 spots are cut out of 1664. This process assumes that strong outliers are caused by defective regions (containing hot pixels and hot ring structures) on the detector that were missed by our bad pixel map. Ultimately we will want to follow up and determine whether spots in these regions can be cut based on more robust criteria. For the present, we focus on contributions from the central part of the distribution.

We found that the correlation functions could be significantly reduced by fitting the ellipticity maps with 2$^{nd}$ order polynomials and computing correlation functions for the residuals (see Figure 15). The model is given by

$$\varepsilon_i^{\text{fit}}(x,y) = a_i + a_{xi}x + a_{yi}y + b_{xi}x^2 + b_{yi}y^2 + b_{xyi}xy \qquad \text{Equation 14}$$

where (x,y) are mask coordinates and *i* denotes the ellipticity polarization. This fit removes large-scale coherent patterns in the PSF that could arise from optical or mask effects that were not eliminated by our focus method. The fit may also be removing detector effects that we are interested in eventually characterizing. However, this process mimics what is done in real weak lensing analyses, where the total PSF (including detector effects) is interpolated between stars and removed from galaxies using much higher order polynomials (or a similar basis of functions).[54,55] Obtaining a small residual correlation function after subtracting only a low order polynomial demonstrates that the bias in the weak lensing signal by the detector is very easily manageable.

The correlation functions for residuals of the polynomial fits are shown in Figure 16 and Figure 17 for coordinate dependent and independent bases, respectively. The residual correlations are relatively stable from run-to-run and consistent with zero over most of the available separation range. The source of the non-zero correlations is not yet clear; however, assuming that they are entirely due to the detector puts an upper limit of $O(10^{-6})$ on detector-induced bias. That bias is two orders of magnitude smaller than the expected weak lensing signal, which is of $O(10^{-4})$. In practice, these correlated errors would be further reduced by fitting more complicated functions. As our experiments continue, we will determine how much of this bias is due to the detector alone and how complex a fit is actually needed to keep the residuals below the error budget.

---

[54] e.g. Rowe, B.T.P. 2010, MNRAS, 404, 350
[55] e.g. Kitching, T.D., Rowe, B., Gill, M., et al., 2013, ApJS, 205, 2, 12, 11

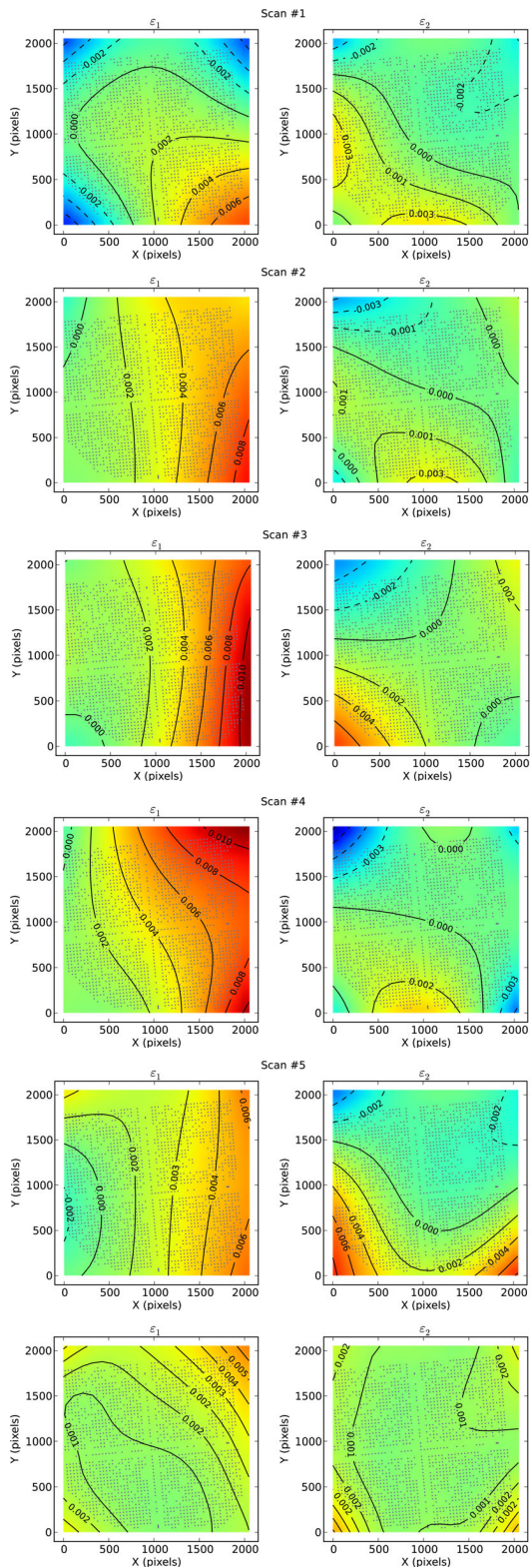

**Figure 15:** Plots of the 2$^{nd}$ order polynomial fits to the best focus ellipticity maps for the five data sets. Maps of $\varepsilon_1$ are on the left, while those for $\varepsilon_2$ are on the right of each figure pair. The standard deviations of the five data sets are shown in the bottom two plots.

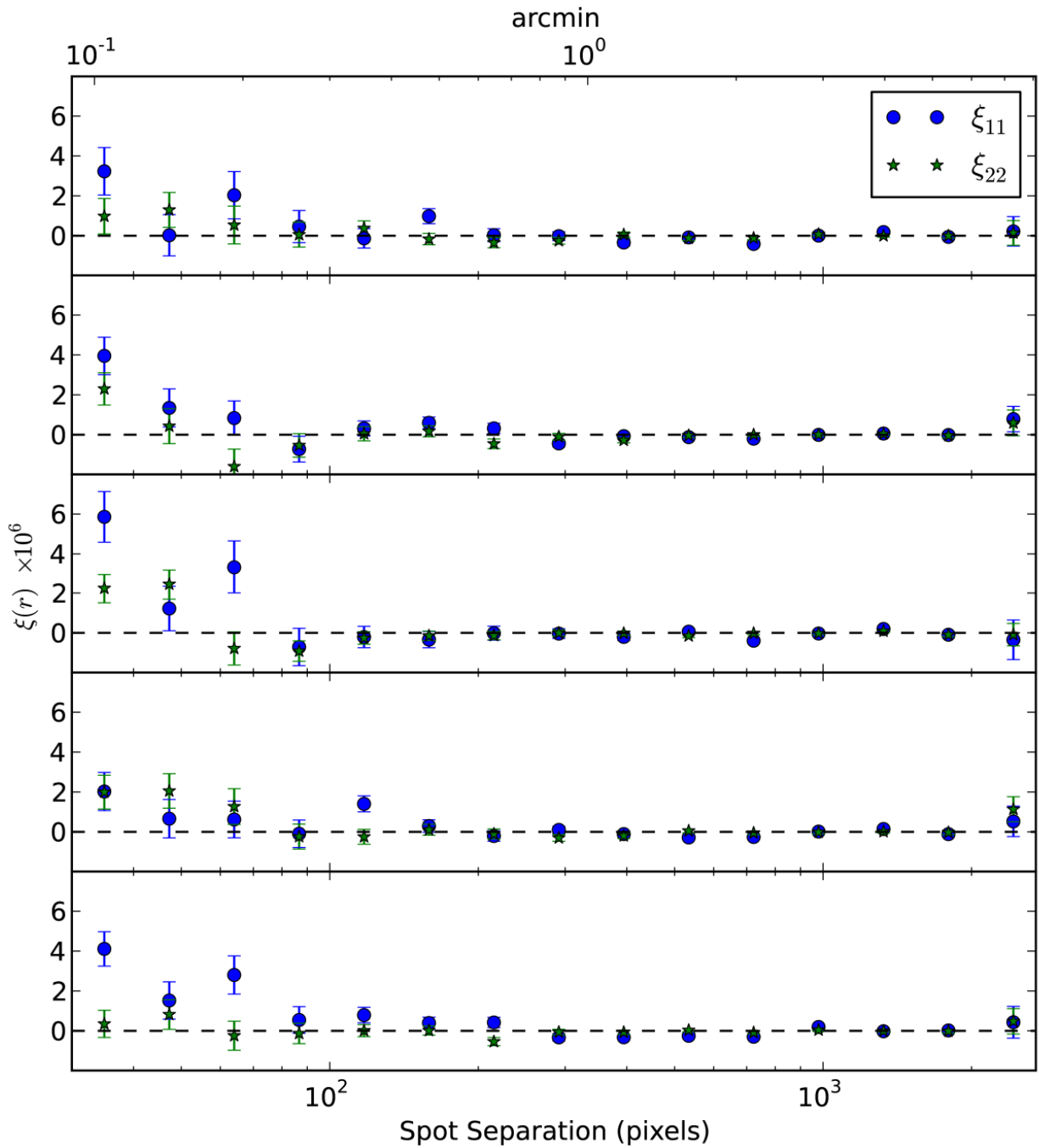

**Figure 16:** Residual correlations after removing fitted 2nd order polynomials from the ellipticity maps. Correlations are calculated in mask coordinates, using Eqn. 9, but we plot separations in units of detector pixels (18µm/pixel). The upper axis along the abscissa has been scaled to 0.18 arcminutes/pixel, which is the proposed scale for WFIRST DRM1. Each panel is for an independent data set.

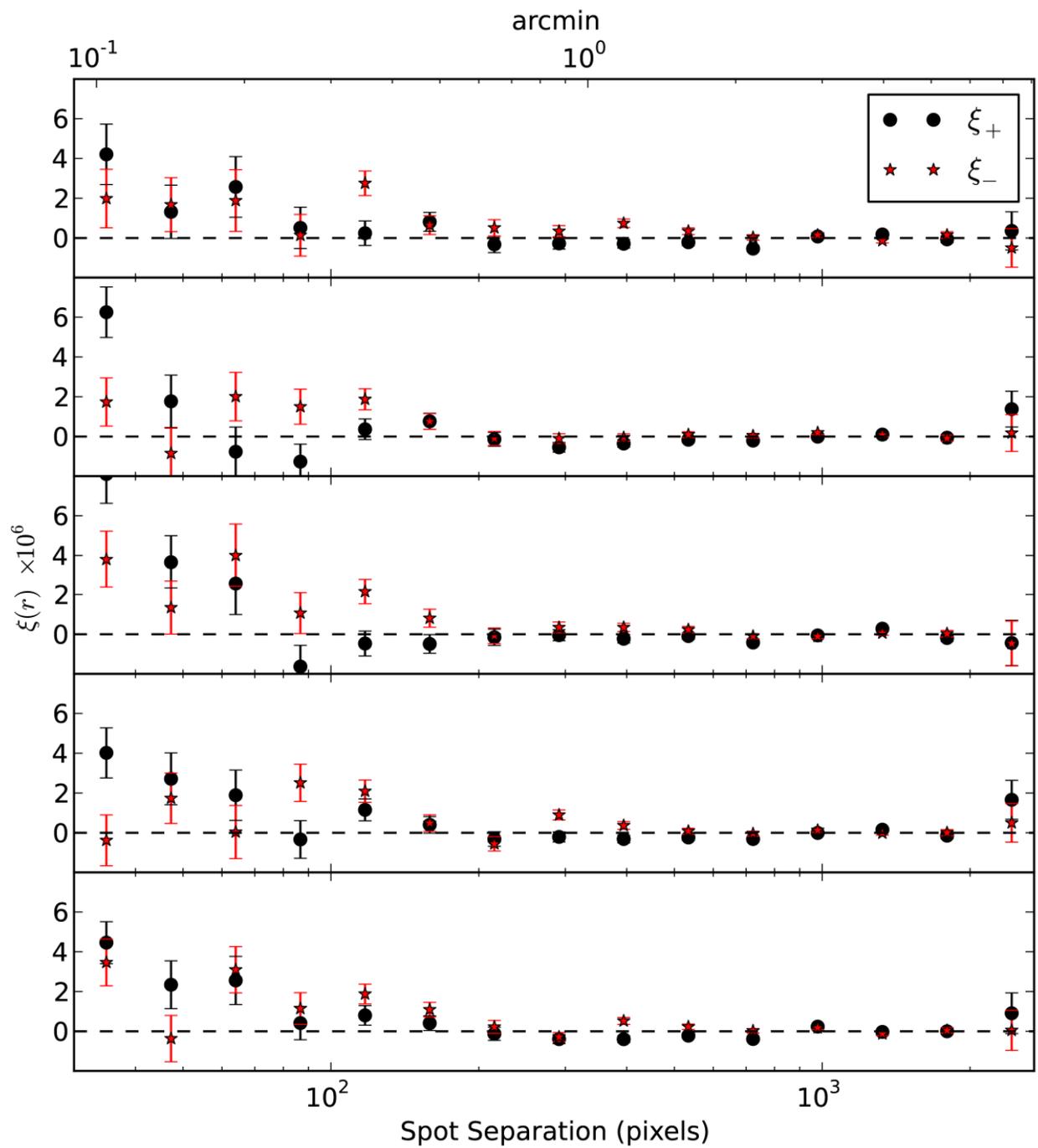

**Figure 17:** Same as Figure 16 but for the coordinate-independent correlation functions defined in Eqn. 12.

# 6. DISCUSSION

## 6.1. *Relevance to weak lensing science*

Weak gravitational lensing observations are a key science driver for WFIRST, which would image the NIR sky using Teledyne HgCdTe detectors hybridized to the Hawaii-xRG (HxRG) family of readout multiplexors. We have performed the first laboratory emulation of weak lensing measurements in order to quantify shape measurement errors induced by an H2RG detector. We focused arrays of 3µm pinhole spots (emulating stellar point sources) across the detector area using our custom precision projector system. We reconstructed oversampled images of each spot from dithered, undersampled images using the IMCOM algorithm. We then measured spot ellipticities to produce maps analogous to shear maps created in weak lensing analyses. After fitting 2$^{nd}$ order polynomials to these maps, we found that the shear correlation functions of the residuals were O($10^{-7}$) in five independent data sets. This result places a conservative upper limit on the spurious correlation functions that would be induced by the detector in a real weak lensing analysis. Our experiment is being improved in order to measure a smaller upper limit and isolate detector-induced errors in the shear correlation functions.

The WFIRST DRM1 would use an H2RG detector like ours with a plate scale of 0.18 arcminutes/pixel, and the upper axes of Figure 16 and Figure 17 have been scaled to reflect this. Any bias to the shear correlation functions must be no more than O($10^{-7}$) in order to avoid biasing an analysis of dark energy parameters; however, any such analysis will be restricted to scales larger than 0.5 arcminutes (or multipoles $\ell < 20{,}000$).[15] Shear correlations cannot be accurately predicted below this scale due to non-linear gravitational clustering and the effects of baryonic matter.[56] Restricted to the scales of interest, our upper limit on the detector-induced correlation functions is in fact $O(10^{-7})$, which is very promising.

There are several differences between our initial experimental setup and the operating configuration of WFIRST. WFIRST detectors would have a 2.4μm cutoff and use either the H2RG (2k x 2k pixels, 18μm pitch) or H4RG format (4k x 4k pixels, 10μm pitch). Our engineering-grade detector is an H2RG with a 1.7μm cutoff. Also, for practical reasons (see section 2.3), we illuminated our source mask with a 0.88μm light source whereas WFIRST will take broadband images at longer wavelengths.[57] For instance, WFIRST-AFTA will image in three bands: J (1.13-1.45), H (1.38-1.77), WFIRST-AFTA F184 (1.68-2.00). These differences could be important if there are wavelength-dependent effects on pixel response. Future iterations of our experiment will determine whether that is the case; however, our initial results are encouraging and suggest that detector-induced shape measurement errors in WFIRST will be manageable.

An important feature of our emulation is the degree of undersampling (or aliasing; see section 2.2). The more undersampled an image is, the smaller the PSF will be relative to a pixel, and the greater effect detector errors will have on shape measurement. Our setup produced images that were slightly more undersampled (Q=1.07) than what is expected for the H band in any WFIRST design reference mission.

---

[56] Semboloni, E., Hoekstra, H., Schaye, J., et al. 2011, MNRAS, 417, 3, 2020
[57] WFIRST-DRM1 bands in μm are J (1.16-1.52), H (1.45-1.91), K (1.83-2.4).

## 6.2. *Future Plans*

Our efforts to date are only the first steps toward fully validating the capability of NIR sensors for the WFIRST mission. Our future plans fall into three main categories:

1. Reducing systematic effects through analysis improvements and updating hardware
2. Running experiments to quantify optical aberrations in the projector and separate them from detector effects
3. Generating data and designing analyses that more closely emulate WFIRST

These efforts will move us toward our ultimate goal of partitioning detector effects on shape measurements and quantifying their impact on weak lensing analyses.

Our first priority is to discover the source of the large-scale run-to-run variability seen in focus positions (section 4.2) and shape measurements (section 5.2). Our working hypothesis is that these fluctuations are the result of varying thermal expansion in the optical system. We are installing an active thermal control system on our optical bench and projector components that will maintain them at a fixed temperature (biased above ambient). In conjunction with the projector's thermally isolating enclosure (panels consisting of foam-core board), this will reduce variations due to thermal expansion and contraction. Image motion must also be improved. To reduce low frequency vibrations, we are testing a tip/tilt stabilization system that will operate at 40Hz. These upgrades should improve the stability of our data and allow us to better probe the sub-pixel response of the detector.

Measuring and modeling aberrations in our optical system will enable us to distinguish them from detector errors. We are quantifying optical aberrations through extra-focal imaging experiments.[58] These experiments use highly defocussed images to measure the Zernike coefficients contributing to wave-front errors across the focal plane. We are also developing analyses to partition errors by separately rotating the mask, detector, and pupil stop about the optical axis. Clearly, shape measurement errors due to the detector should not change as the mask and pupil stop are rotated. Measuring shapes generated by elliptical pupils or mask sources also allows useful consistency checks. For instance, we will check that an elliptical pupil adds a uniform ellipticity vector to all objects in the field, that the vector changes sign under 90 degree rotations, and that its magnitude is invariant under all rotations.

In this study, sensor calibrations have only included subtraction of matched-exposure darks (capturing dark current and electrical offsets) and flat fielding (correcting both gain and sensitivity). In future, corrections for linearity, interpixel crosstalk anisotropy and image persistence may be applied to test their impact in the shape measurement noise and systematics. It is also clear in several figures that our target mask does not cover the entire detector with sources at once. Future experiments will sample more of the detector surface, particularly near the edges.

There are various ways to bring our emulations more in line with expected WFIRST weak lensing data analysis techniques and data acquisition. It will be necessary to switch our illumination to longer wavelength LEDs that match the WFIRST filter set (if wavelength dependent detector effects are significant). We also plan to repeat our tests over the range of flux levels expected from WFIRST sources. Another important change will be to replace our unresolved point sources with extended objects with intrinsic ellipticities, perhaps approximating galaxy light profiles. We will want to investigate how shape measurement errors depend on the size and orientation of extended objects. Furthermore, we plan to match the WFIRST dithering strategies. For our correlation function measurements, each spot image was reconstructed from 15 dithered exposures in order to boost the signal to noise ratio. WFIRST will use 5 to 9 dithers per image depending on the mission design and filter. Finally, we will repeat our emulations on an H4RG detector to ensure that it behaves similarly to the H2RG.

---

[58] Roodman, A. 2010, Proc. SPIE, 7735, 77353T

## 7. ACKNOWLEDGMENTS

We thank E. Jullo of Laboratoire d'Astrophysique de Marseille and V. Velur for their contributions to the analysis pipeline and optical design and Richard Massey of the Department of Physics at Durham University for many useful discussions regarding the projector design. We also thank C. S. Peay, P. G. Ringold and C. J. Wrigley of JPL, as well as, D. Hale and K. Bui of Caltech Optical Observatories for their technical support. Finally, we thank our anonymous referee for improving the quality and clarity of this manuscript.

This research was carried out at the Jet Propulsion Laboratory and California Institute of Technology, under a contract with the National Aeronautics and Space Administration. We are grateful to the following organizations and programs for their support of this effort: internal JPL Research and Technology Development (RTD) and Director's Research Development Fund (DRDF) programs; US Department of Energy's (DOE) Supernova Acceleration Probe (SNAP) and Joint Dark Energy Mission (JDEM) projects; the NASA Wide Field IR Survey Telescope (WFIRST) and Joint Dark Energy Mission (JDEM) project offices. CS acknowledges support from the NASA Postdoctoral Program Fellowship Program from Oak Ridge Associated Universities. BR acknowledges support from European Research Council in the form of a Starting Grant with number 240672.